\documentclass[]{aastex631}

\usepackage{amsmath}
\usepackage{longtable}
\usepackage{tablefootnote}
\usepackage{cancel} 
\usepackage{hyperref}

\begin{document}

\title{CO Observations of the SMC-N66 H{\sc ii} Region with ALMA: Properties of Clumps along Filamentary Molecular Clouds and Possible Expansion Motion }

\email{naslim.n@uaeu.ac.ae}
\author{Batool Ilyasi}
\affiliation{Department of Physics, College of Science, United Arab Emirates University (UAEU), Al-Ain, UAE, 15551}

\author[0000-0001-8901-7287]{Naslim Neelamkodan}
\affiliation{Department of Physics, College of Science, United Arab Emirates University (UAEU), Al-Ain, UAE, 15551}

\author[0000-0002-2062-1600]{Kazuki Tokuda}
\affiliation{Department of Earth and Planetary Sciences, Faculty of Sciences, Kyushu University, Nishi-ku, Fukuoka 819-0395, Japan}
\affiliation{National Astronomical Observatory of Japan, National Institutes of Natural Science, 2-21-1 Osawa, Mitaka, Tokyo 181-8588, Japan}
\affiliation{Department of Physics, Graduate School of Science, Osaka Metropolitan University, 1-1 Gakuen-cho, Naka-ku, Sakai, Osaka 599-8531,11 Japan}

\author{Susmita Barman}
\affiliation{Department of Physics, College of Science, United Arab Emirates University (UAEU), Al-Ain, UAE, 15551}
\affiliation{School of Physics, University of Hyderabad, Prof. C. R. Rao Road, Gachibowli, Telangana, Hyderabad, 500046, India}

\author[0000-0003-2248-6032]{ Marta Sewilo}
\affiliation{ Exoplanets and Stellar Astrophysics Laboratory, NASA Goddard Space Flight Center, Greenbelt, MD 20771, USA}

\affiliation{Department of Astronomy, University of Maryland, College Park, MD 20742, USA}

\affiliation{Center for Research and Exploration in Space Science and Technology, NASA Goddard Space Flight Center, Greenbelt, MD 20771, USA}

\author[0000-0003-2062-5692]{Hidetoshi Sano}
\affiliation{Faculty of Engineering, Gifu University, 1-1 Yanagido, Gifu 501-1193, Japan}
\affiliation{Center for Space Research and Utilization Promotion (c-SRUP), Gifu University, 1-1 Yanagido, Gifu 501-1193, Japan}

\author{Toshikazu Onishi}
\affiliation{Department of Physics, Graduate School of Science, Osaka Metropolitan University, 1-1 Gakuen-cho, Naka-ku, Sakai, Osaka 599-8531,11 Japan}



\begin{abstract}

The star-forming region N66, as a host of the majority of OB stars in the Small Magellanic Cloud, provides a unique opportunity to enhance our understanding of the triggers of high-mass star formation. We investigate the properties of the molecular cloud in N66 using the \textsuperscript{12}CO(1-0) data obtained with the Atacama Large Millimeter/submillimeter Array.  A cloud decomposition analysis identified 165 independent cloud structures and substructures. The size-linewidth scaling relation for the entire region exhibits an index of 0.49, indicating that the region is in a state of virial equilibrium. In contrast, a detailed analysis of the central N66 region revealed a size-linewidth scaling relation with an index of 0.75, suggesting that distinct factors are influencing the dynamics of this central area. Averaging the spectra in the central N66 region revealed three distinct velocity peaks at 145, 152, and 160 $\mathrm{km \, s^{-1}}$, indicating that some kinds of interactions are occurring within the cloud. The analysis of the position-velocity diagrams in the central region revealed a ring-like structure, indicating the presence of an expanding bubble.
 The bubble exhibits supersonic characteristics, with an expansion velocity of $v_{\mathrm{exp}} \approx 11$ $\mathrm{km \, s^{-1}}$, and an overall systemic velocity of $v_{\mathrm{sys}}\approx $ 151 $\mathrm{km \, s^{-1}}$. The radius is estimated to be in the range of  $r \approx [9.8 - 12.9] \pm 0.5$ pc and is approximately 1.2 Myr old.

\end{abstract}
\keywords{ISM: bubbles -- Stars: formation -- Small Magellanic Cloud --  ISM: kinematics and dynamics}


\section{Introduction} \label{sec:intro}

Molecular clouds are the coldest and densest components of the interstellar medium (ISM), characterized by sub-parsec-sized clumps where star formation takes place. These clouds are predominantly composed of molecular hydrogen (H$_2$), which remains largely undetectable through direct observational techniques. Consequently, the properties of molecular clouds are inferred from observations of tracers, such as carbon monoxide (CO) and its isotopes (\textsuperscript{12}CO, \textsuperscript{13}CO, and C\textsuperscript{18}O) \citep{Dame87, Mizuno95, Kawamura98, Onishi96}, and  [C{\sc i}] and [C{\sc ii}] that trace the ``CO dark" H$_2$ gas \citep{Papadopoulos04, Israel16, Jameson18} and is useful in the low metallicity environments. 



 The molecular clouds in the Milky Way (MW) exhibit a size-linewidth scaling relation indicating that the cloud is virialized \citep{Solomon87, Larson81}. These observations have been conducted not only in the MW galaxy but also in several other galaxies within the Local Group \citep{Kawamura98, Fukui08, Bally80, Ohno23}, with analyses that provide valuable information on the statistical characteristics of the molecular clouds.

The Small Magellanic Cloud (SMC) is a great source to study star formation and evolution at low metallicity (0.2 Z$\odot)$ \citep{Russell92, Rolleston03, Lee05,Hunter07, Tokuda21}, since it is only $ \approx 62$ kpc \citep{Graczyk20} away. It has a CO mass spectrum that follows a power index of $\alpha \approx 1.7$ \citep{Takekoshi17, Ohno23}, similar to that of the MW ($\sim 1.8$;  \citealt{Heyer01}) and the Large Magellanic Cloud (LMC) ($\sim 1.8$; \citealt{Fukui08}). Moreover, the CO clouds in the SMC follow a size-linewidth relation that is a factor of about 1.5 smaller than that of the MW \citep{Ohno23,Saldano23}, indicating that either the virial equilibrium conditions in the SMC are at a lower column density than that of the MW or that the clouds are unstable against the free fall collapse.

\begin{figure*}
    \centering
    \includegraphics [width=10 cm]
     {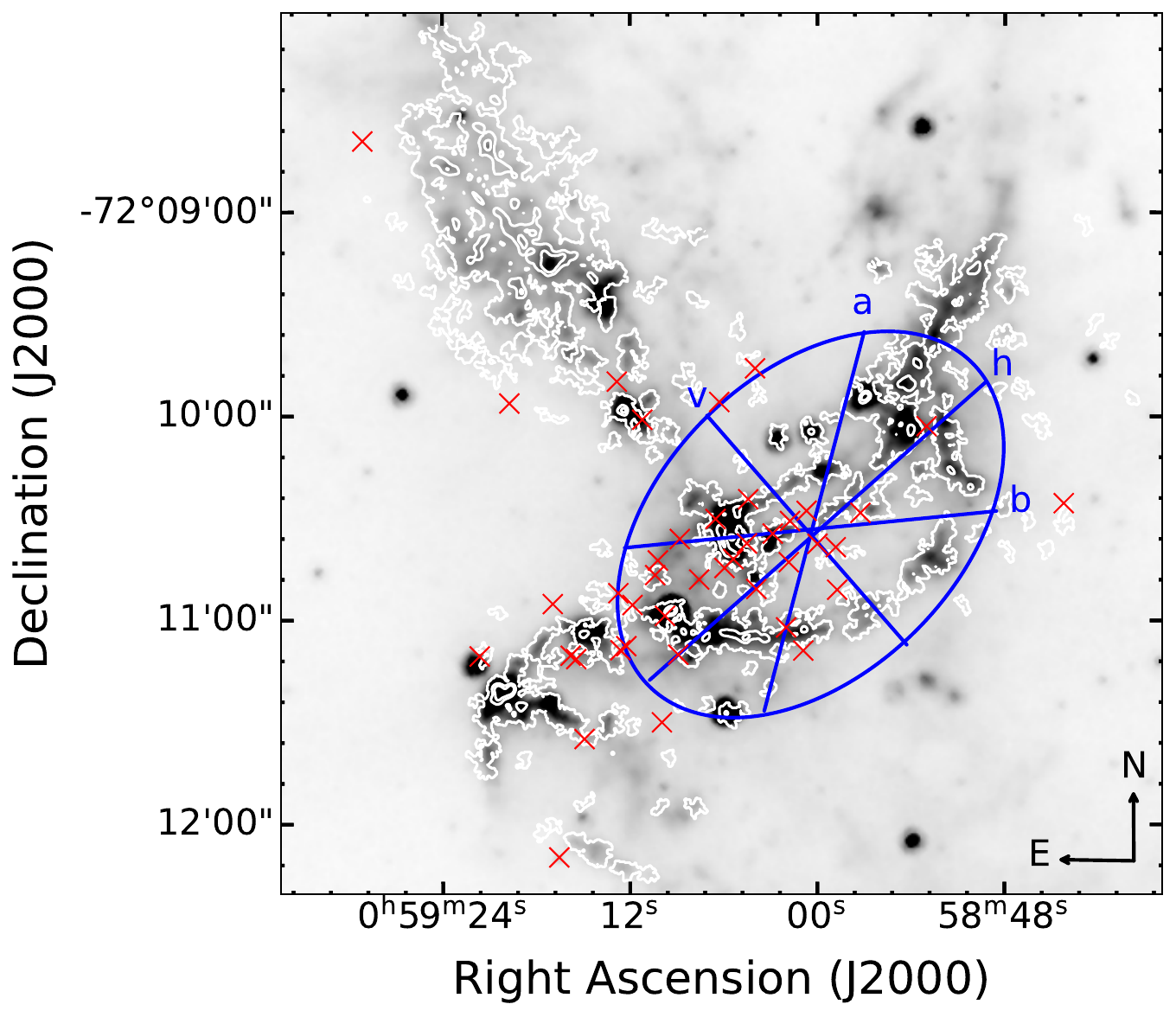} 
    \caption{ The structure of N66 in a Spitzer $8.0 \, \mu m $ map  \citep{Gordon11}, which traces the PAH emission, is shown alongside ALMA's \textsuperscript{12}CO(1–0) integrated intensity contours at the level: 0, 20, 40, and 80 K km s$^{-1}$. The OB stars \citep{Dufton19} are indicated in red crosses (x). The blue elliptical shape indicates
the central region that is investigated for expansion. With the: v, h, a, and b lines indicating the locations at which the PV cuts are taken. 
}
    \label{fig:IRAC+YSO+stars}
\end{figure*}

The nebula, N66, is the largest and most luminous H{\sc ii} region in the SMC \citep{Henize56}, with the largest star formation rate. It is associated with NGC 346, the stellar cluster located within the nebula, which has a great variety of stellar populations, including young stellar objects (YSOs; \citealt{Simon07,Sewilo13}), pre-main-sequence (PMS; \citealt{Nota06,Hennekemper08}), and main sequence stars \citep{Massey89, Evans06,Dufton19}. It contains more than 300 OB stars \citep{Dufton19}, which makes up the majority of the known hot star population in the SMC \citep{Massey89}. Most of those massive stars are at the central bar of the nebula embedded within the champagne-shaped structure seen in Figure \ref{fig:IRAC+YSO+stars}.

Massive stars play a key role in the change within the physical and dynamic condition of the ISM they are embedded within. These massive stars are a major source of photoionization, stellar wind, and shock, which in return can be the trigger of star formation in the region \citep{Elmegreen77}. Since N66 still contains some of its residual gas, this suggests that a supernova explosion from its massive stars has yet to occur in the central region \citep{Naze02}. Due to the SMC's lower metallicity, it has been found that the stellar winds in N66 are not powerful enough to sweep the residual gas and that the ISM in N66 is rather shaped by stellar radiation and photoionization \citep{smith08}.

 The study by \citet{Neelamkodan21} highlights the role of cloud-cloud collision in triggering star formation in the northern N66 region. However, it's crucial to recognize that N66 hosts multiple PMS clusters with a wide age range, as noted by \citet{Hennekemper08}. The massive stars at the center of the system are likely formed several million years after those in the outer regions, as discussed in \citet{Dufton19} and \citet{Cignoni11}. This suggests that the star formation mechanisms for the northern part of N66 differ from those of the central bar population, indicating a complex history of star formation in this region \citep{Hennekemper08}.

An expanding ring of shocked molecular gas has been found in N66 \citep{Danforth03,Gouliermis08}, using [C{\sc ii}] \citep{Israel16}, at the center of the stellar cluster, with the central bar population being reported to have a clear rotational signature \citep{Sabbi22, Zeidler22}. This spiraling motion is suspected to be triggering the more recent episode of star formation \citep{Sabbi22}. The expanding bubble of the ionized gas in the N66 central region might be caused and driven by the stellar winds and ionizing fluxes of the massive stars in the region \citep{Zeidler22}. 

Photoionization from massive stars can play a crucial role in the evolution of a galaxy by either triggering or suppressing star formation \citep{ Elmegreen77, Efstathiou92}. Such feedback in the star formation regions can drive the formation of bubbles within them \citep{Collins22}.

This paper examines the molecular clump properties of N66, the most luminous region in the SMC, using Atacama Large Millimeter/submillimeter Array (ALMA) observations of the \textsuperscript{12}CO(1–0), with the aim to determine what makes it unique in comparison with the whole SMC. We describe the ALMA observation and data analysis in Section \ref{sec:obs}. We present and analyze the molecular cloud and clump properties such as size, velocity dispersion, mass, luminosity, and their association with star formation in Section \ref{sec:anays}. We examine the scaling relation of N66 and compare it with that of the SMC, MW, and the LMC clouds in Section \ref{sec:scaling-rel} to determine whether the power law relation holds in such an environment. In Section \ref{sec:massspectra}, we investigate the cloud mass spectrum and see if that for N66 specifically aligns with that of the SMC. In Section \ref{ssec:Kinematics}, we study the kinematics of the clouds in the region and investigate whether the expansion movement of the stars at the central region is also occurring at the molecular scale. We geometrically analyze the expansion bubble and identify its velocity parameters and age approximation in Section \ref{sec:expansion}. In Section \ref{sec:discu}, we discuss the meaning behind the scaling relation (Section \ref{sec:disc-scaling}) and the $X_{\mathrm{CO}}$ factor (Section \ref{sec:disx_X}). 
Moreover, in Section \ref{sec:drivesit}, we investigate the driving force of the expansion and if the photoionization of the stars is a significant trigger of it. Lastly, we summarize our results in Section \ref{sec:sum}.

\section{Observations} \label{sec:obs}

We use the archival ALMA Band 3 (3 mm) data for N66 (P.I., Erik Muller, $ 2015.1.01296.$S). The 12m-Array data were obtained in ALMA Cycle 3 and 4 with the array in C36-1/2, C36-2/3, and C40-6 configurations. The observations targeted the $^{12}$CO(J = 1–0) line, with a rest frequency of 115.27 GHz, with 3840 channels. It had a resolution of 77 kHz for the frequency resolution and a $0.2 $ km s$^{-1}$ for the velocity resolution. 
The image processing and deconvolution followed the same steps as \citet{Neelamkodan21}, which were performed using the Common Astronomy Software Application (CASA; \citealt{McMullin07}). The ``Briggs" weighting system was used for the \textsc{tclean}, with a 0.5 robust parameter. The dirty and residual images were automatically chosen using the \textsc{auto-multithresh} process \citep{Kepley20} for \textsc{tclean} until reaching the $\sim 1\sigma$ noise level.
The resultant beam size was $2.^{\prime \prime} 0 \times 1.^{\prime \prime}8$ (P.A. = $10^\circ.0 ; 0.58 \times 0.52$ pc$^2$) with a $0.022$ Jy beam$^{-1} (=0.59$ K) noise level.

\section{Analysis} \label{sec:anays}

\subsection{Morphology} \label{sec:Morph}

The ALMA \textsuperscript{12}CO(1–0) observations of N66 reveal a clumpy molecular gas, as shown in Figure \ref{fig:IRAC+YSO+stars}, especially at the sub-parsec scales, with clear filamentary structure at parsec scales. The filaments in the central region of N66 exhibit a semicircular structure that is consistent with the ``champagne" flow model. 
The presence of an ionizing front within a strong density gradient in a molecular cloud causes the inner regions of the cloud to expand faster than the outer regions. This leads to a shock that passes through the cloud and accelerates the gas to supersonic velocities. This is known as the ``champagne" flow model \citep{Tenorio79, Franco90}. 

\begin{figure}
    \centering
     {{\includegraphics[width=17.5 cm]
   {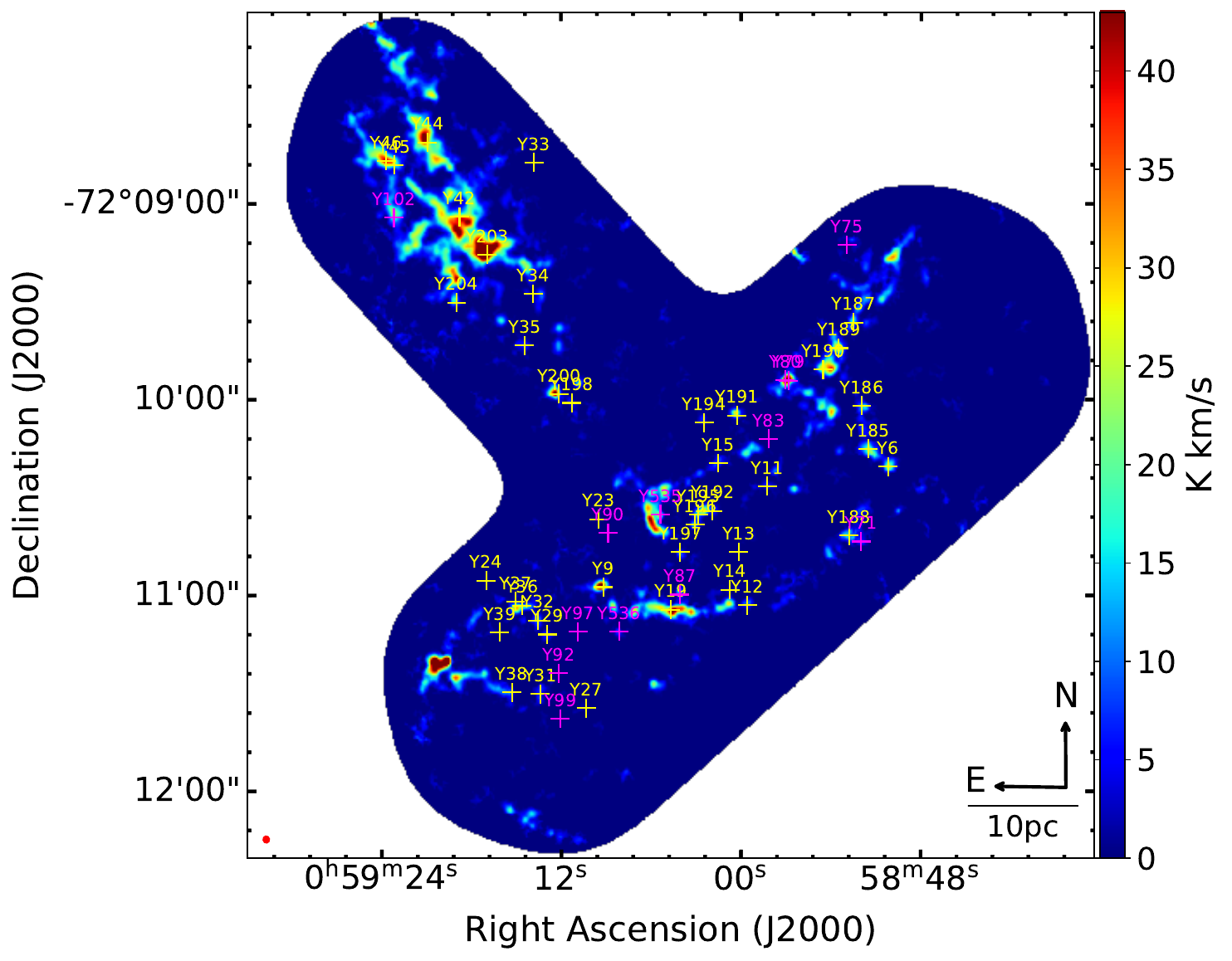} }}%
    \qquad
    \caption{The \textsuperscript{12}CO(1–0) integrated intensity map of N66 is shown along with the highly reliable YSOs that are labeled in yellow, and the less reliable YSOs labeled in magenta \citep{Sewilo13,Bolatto07,Simon07}. The size of the ALMA beam ($2.^{\prime \prime} 0 \times 1.^{\prime \prime}8$) is shown as a red ellipse in the left bottom corner. }
    \label{fig:M0_n66}
\end{figure}

The semicircular pattern seen in N66 probably results from the projection of a supersonic bubble surrounded by a champagne flow \citep{Ye91}. This supersonic bubble region is indicated by a blue ellipse in Figure \ref{fig:IRAC+YSO+stars} and is referred to as the expansion region. The curved filaments surrounding the bubble contain numerous clumps and cores (clumps are dense sub-parsec structures in the molecular cloud, and cores are denser sub-clumps structures within the clump). These clumps and cores are associated with YSOs, shown in Figure \ref{fig:M0_n66}. The filamentary structures traced by \textsuperscript{12}CO(1–0) align with that of the PAHs traced by the IRAC $8.0 \, \mu \mathrm{m}$ emission \citep{Gordon11}. These structures, with their clumps and cores, exhibit a unique and complex nature. Their properties are identified using the dendrogram analysis.

\subsection{Clump Identification} \label{sec:clump_id}

The molecular clouds' structure follows a hierarchy, with the highest density molecular cores occupying the smallest volumes and are embedded within the larger, less dense part of the cloud \citep{Lada92}. In this study, \textsc{Astrodendro} \citep{Rosolowsky08}  was used to identify the hierarchy of the \textsuperscript{12}CO(1–0) emission in N66. \textsc{Astrodendro} is a Python package that divides the three-dimensional astronomical data into a structure tree \citep{Houlahan92, Rosolowsky08}. The smallest and brightest portion of the three-dimensional contours is considered a leaf. Those leaves are embedded within the larger less bright portion of the contours, and together they are considered as trunks. Leaves have sizes that are usually within the sub-parsec scale, while the trunks are more of a parsec scale structure. For this study, dendrogram leaves represent the densest structures or molecular cores, whereas the trunks represent the molecular clumps. This molecular cloud division and characterization technique provides the tool to measure different properties of the molecular cloud, such as velocity, size, virial mass, and luminosity. 


In a three-dimensional data cube, \textsc{Astrodendro} finds distinct isosurfaces from each emission region and uses the moment of volume-weighted intensities from each pixel to identify the characteristics and properties \citep{Rosolowsky08}.

The parameters used to extract the clumps and leaves were chosen as: \textsc{min{\_}value} is 5 times that of the noise level, \textsc{min{\_}delta} is 3 times that of the noise level, and \textsc{min{\_}npix} is 3 times that of the beam size. The computed trunks can be seen in figure \ref{fig:Moment1}.

The parameters identified include the X and Y positional dimensions, velocity dimension, velocity dispersion ($\sigma_{\mathrm{v}}$), rms sizes of the major ($\sigma_{\mathrm{major}}$) and minor ($\sigma_{\mathrm{minor}}$) axes, the rms of the clumps radius ($\sigma_{\mathrm{r}}$), the angle of the major axis, and the integrated flux density. The radius of the spherical cloud is acquired using the radius rms by applying $ R \approx 1.91 \sigma_{\mathrm{r}}$ \citep{Solomon87, Rosolowsky06}.

We identified 165 significant structures: 104 were molecular clumps (shown in Table \ref{tab:Trunks}), and 61 were leaves or molecular cores (shown in Table \ref{tab:leaves}). The statistical summaries of the clumps and cores are shown in Table \ref{tab:clump_stat_sum} and \ref{tab:cores_stat_sum} respectively. 

\begin{longtable}{lllllllllllll}
\caption{\textsuperscript{12}CO(1–0)  Cores (leaves) statistical summary. } \\

\hline
   &  Min & Max   & Mean   & Median  \\ 
\hline
\endfirsthead 
  $v_{\mathrm{cen}}$ (km s$^{-1}$)  &  139	&164 &	157 &	159 \\
  $\sigma_{\mathrm{maj}}$ (arcsec)   &1.19&	4.29	&2.15	&1.94\\
  $\sigma_{\mathrm{min}}$ (arcsec)  & 0.50&	1.94&	0.98	&0.90\\
  $v_{\mathrm{rms}}$ (km s$^{-1}$)  & 0.23	&1.19	&0.61&	0.59\\
  $L_{\mathrm{CO}}$(K km s$^{-1}$ pc$^2$)  & 1.36&	189.3 &	25.7 &	12.4 \\
  Radius (pc) & 0.55 &	1.27&	0.79&	0.73\\
  $M_{\mathrm{vir}}$ ($\mathrm{M_{\odot}}$) & 32&	1530&	367&	272\\
  \hline
\label{tab:cores_stat_sum}
\end{longtable}

\begin{longtable}{lllllllllllll}
\caption{\textsuperscript{12}CO(1–0)  Clumps (trunks) statistical summary. } \\
\hline
   &  Min & Max   & Mean   & Median  \\ 
\hline
\endfirsthead 
  $v_{\mathrm{cen}}$ (km s$^{-1}$)  &  124	&	168	&	150	&	152	\\
  $\sigma_{\mathrm{maj}}$ (arcsec)  & 1.03	&	21.29	&	2.56	&	1.85	\\
  $\sigma_{\mathrm{min}}$(arcsec)  & 0.59	&	8.68	&	1.20	&	0.99	\\
  $v_{\mathrm{rms}}$ (km s$^{-1}$)  &0.32	&	2.35	&	0.79	&	0.70	\\
  $L_{\mathrm{CO}}$ (K km s$^{-1}$ pc$^2$)   & 0.71	&	2346	&	43.39	&	3.04	\\
  Radius (pc) & 0.55	&	7.53	&	0.95	&	0.75	\\
  $M_{\mathrm{vir}}$ ($\mathrm{M_{\odot}}$) & 69	&	32212	&	1278	&	376	\\
  \hline

\label{tab:clump_stat_sum}

\end{longtable}



\subsection{ CO Molecular Mass And Luminosity  } 

Assuming that N66 is virialized, we are able to calculate the virial mass of the \textsuperscript{12}CO(1–0) cores and clumps using   \citet{Wong11} virial mass approximation:
\begin{equation}
M_{\mathrm{vir}}[M_{\mathrm{\odot}}] = 1040 \, \sigma_{\mathrm{v}} \, ^2  R 
\end{equation}

which assumes that the molecular clouds are spherical with truncated density profiles
$\rho \propto  r^{-1} $ ($\rho$ is density and $r$ is radius).

The luminosity of the molecular cloud, $ L_{\mathrm{CO}} $, was obtained using the fluxes from \textsc{Astrodendro} of the cores and the clumps \citep{Rosolowsky06}, where:

\begin{equation}
L_{\mathrm{CO}} \mathrm{[K \, km \, s^{-1} pc^{2} ]} = F_{\mathrm{CO}}(0 \mathrm{K)} \mathrm{[K km s^{-1} arcsec^2 ] (d[pc])^2} \times \left( \frac {\pi}{180 \times 3600} \right) ^2
\end{equation}

Where $F_{\mathrm{CO}}$ is the integrated flux, and d is the distance to N66 in parsec (62 kpc).

Using the velocity dispersion and radius of the clumps and cores, we estimated their virial masses. Additionally, we calculated the CO luminosity of the clumps and cores based on their fluxes (Table \ref{tab:Trunks} and \ref{tab:leaves}).




\section{Results} \label{sec:result}

\subsection{Scaling Relations} \label{sec:scaling-rel}
\subsubsection{The Size-Linewidth Scaling Relations} \label{sec:size-linewidth}

Using the obtained parameters, we were able to examine the properties of the N66 molecular cloud and determine whether it is gravitationally bound or if other forces are influencing its dynamics. To test if the molecular cloud is virialized, we study the size-linewidth relationship of the clumps.  According to \citet{Larson81} size-linewidth scaling relationship, the velocity dispersion is proportional to the radius of the cloud raised to the power of 0.5, 
$\sigma_{\mathrm{v}} \approx 0.72 \times R^{0.5}$ \citep{Solomon87}. There have been observations of various molecular clouds in the MW that demonstrate a power-law relationship between velocity dispersion and size, typically in the range of $ \sigma_{\mathrm{v}} \approx[  R ^{0.2}   - R^{ 0.6}  ] $ \citep{Larson81, Solomon87, Leisawitz90, Caselli95}.

To analyze the size-linewidth relationship of N66, we plotted the velocity dispersion against the radius and performed a power-law fit. This allowed us to determine how well the observed data align with the expected scaling relationship. When investigating the clumps, it gives us a power fit of $ \sigma_{\mathrm{v}}  = 0.85 \times  R^{ 0.54}$.

We see that this molecular cloud overall does indeed follow \citet{Larson81} 0.5 scaling relation. 

\begin{figure}
    \centering
    \includegraphics[width=8.5 cm]
    {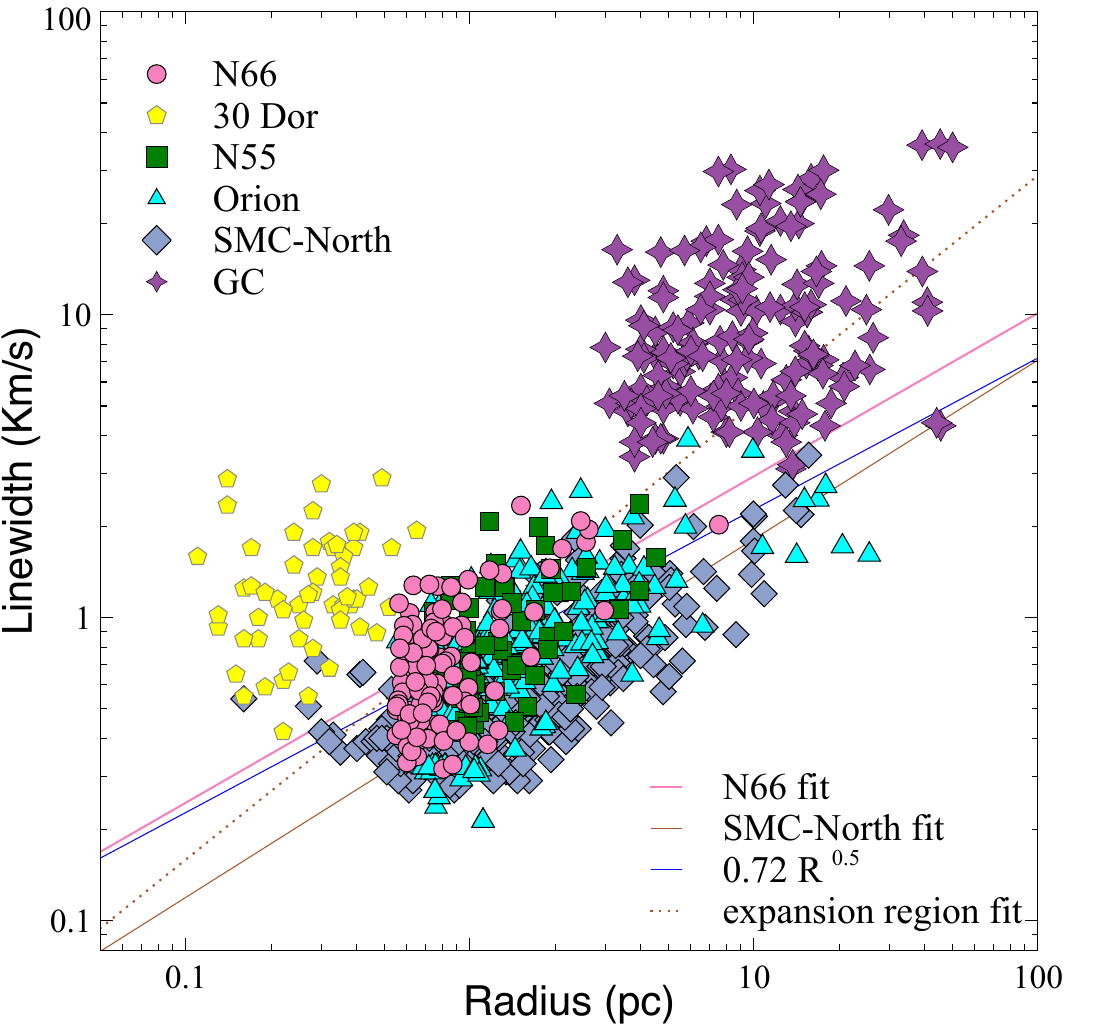}
    \caption{
The size-linewidth scaling relations for N66 clumps using \textsuperscript{12}CO(1–0) emission is shown, in comparison with that of Orion, N55, and Galactic center, as well as the \textsuperscript{12}CO(2-1) transition for 30 Doradus and northern part of the SMC. The N66 clumps as a whole follow a $\sigma_{\mathrm{v}}  \propto  R^{ 0.5}$ power law relation, which agrees well with that of the Orion and N55 clumps. When investigating the clumps in the N66 expansion region alone (indicated by an ellipse in Figure \ref{fig:IRAC+YSO+stars}), the power law relation follows $\sigma_{\mathrm{v}}  \propto  R^{ 0.75}$.}
    \label{fig:Cloud_comparision}
\end{figure}

To better understand where N66 clump results fall, a comparison with other molecular cloud clumps is needed. Figure \ref{fig:Cloud_comparision} show a comparison with the \textsuperscript{12}CO(1–0) clumps of Orion \citep{Dame01}, N55 \citep{Naslim15} and Galactic center \citep{Oka01} clouds alongside the \textsuperscript{12}CO(2-1) clumps of 30 Doradus \citep{Indebetouw13}, and northern part of the SMC \citep{Ohno23}. The data from Orion, N55, and the SMC were decomposed using the same \textsc{Astrodendro} package used for N66.

N66 falls within the size-linewidth relation observed for Orion, N55, and the northern region of the SMC, with the fit exhibiting a power index of comparable values. However, when investigating the intercept of the fit, the northern region in the SMC has an $\approx 0.2$ lower than that of the MW \citep{Ohno23, Saldano23}. The N66 alone has $\approx 0.1$ higher intercept than the MW and $\approx  0.3$ more than the general northern part of the SMC.
 

\subsubsection{Virial Mass–Luminosity Scaling Relations} \label{sec:MvsL_scaling relation}

\begin{figure}
    \centering
    \includegraphics[width=8.5 cm]
    {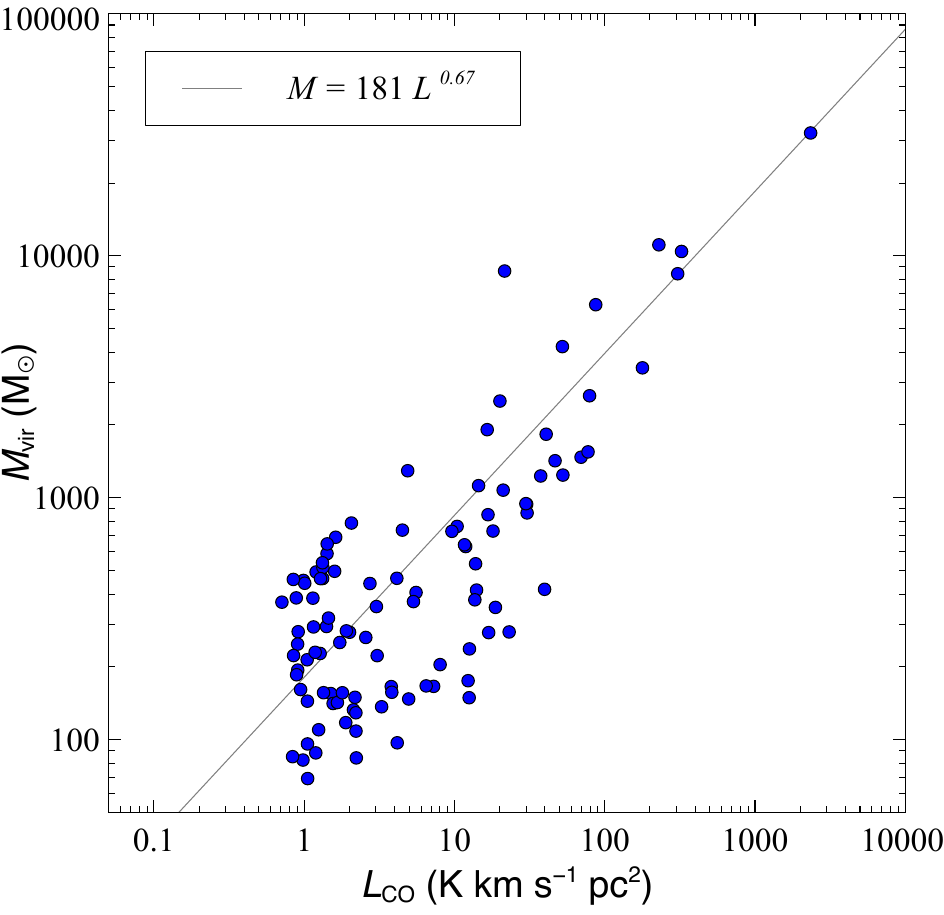}
    \caption{Virial mass vs. luminosity relation for \textsuperscript{12}CO(1–0) clumps in N66. It is
fitted with a power function of $M= 181 \, L^{0.67}$. }
    \label{fig:MvsL}
\end{figure}

Figure \ref{fig:MvsL} shows the virial mass versus luminosity scaling relations for N66 \textsuperscript{12}CO(1-0) clumps. It shows that the N66 follow a power law relation of, $M_{\mathrm{vir}}= 181 \, L ^{0.67} $. 
 The CO-to-H$_2$ conversion factor ($X_{\mathrm{CO}}$), can be estimated using the Equation \citep{Bolatto13}: 

\begin{equation}
 X_{\mathrm{CO}} = \frac{M\textsubscript{vir}} {L_{\mathrm{CO(1-0)}}} \frac{1}{m_{\mathrm{H}} \, \mu} = \frac { \alpha_{\mathrm{CO(1-0)}} } {m_{\mathrm{H}} \, \mu}
 \label{eq:XCO}
\end{equation}

Where $m_{\mathrm{H}}$ is the hydrogen mass, and $\mu$ is the mean
molecular weight of hydrogen (2.7). 

The clumps data was divided into two, those with luminosities less than $4 \, \mathrm{K \, km \,s^{-1} pc^2}$ and those with luminosities greater than that. For luminosities greater than $4 \, \mathrm{K \, km \,s^{-1} pc^2}$, using the power fit obtained and the median luminosities of $17.5 \, \mathrm{K \, km \,s^{-1} pc^2}$ of our data, this results in $\alpha _{\mathrm{CO}} = 70.4 \: \mathrm{ M_{\mathrm{\odot}}  (K \, km \,s^{-1} pc^2)^{-1}}$. Substituting that into Equation \ref{eq:XCO}, the median $X_{\mathrm{CO}}$ of the trunks with $L_{\mathrm{CO}} > 4 \, \mathrm{K \, km \,s^{-1} pc^2}$ in N66 is $ X_{\mathrm{CO}}= 3.26 \times 10 ^{21}\mathrm{ \, cm^{-2} (K \, km \,s^{-1})^{-1} }$. Similarly, for luminosities less than $4 \, \mathrm{K \, km \,s^{-1} pc^2}$, our median luminosities is found to be  $1.34 \, \mathrm{K \, km \,s^{-1} pc^2}$, and $\alpha _{\mathrm{CO}}$ is $ 164.3 \: \mathrm{ M_{\mathrm{\odot}}  (K \, km \,s^{-1} pc^2)^{-1}}$, resulting in  $X_{\mathrm{CO}}= 7.6 \times 10 ^{21}\mathrm{ \, cm^{-2} (K \, km \,s^{-1})^{-1} }$.


\subsection{The CO Cloud Mass Spectrum  } \label{sec:massspectra}

The mass spectrum distribution of the molecular cloud is a method to represent how the molecular cloud transforms into stars, which in turn represents the starting stage of the initial mass function (IMF). This frequency distribution of the mass in the molecular cloud can be represented with different forms \citep{Kawamura98}. Using the differential form of the power law,  $ \frac{dN}{dM} \propto M^{-\alpha} $, where dM represents the size of the mass bin and dN indicates the number of clouds within that mass bin, we can analyze the distribution of clouds in relation to their masses. This differential representation is sensitive to the mass bin chosen and can be affected by it. On the other hand, the cumulative mass distribution(CMD) doesn't have that issue, and its power law is often used to represent the mass spectrum. With,   

\begin{equation}
 N( > M) \propto M^{-(\alpha -1) }
\end{equation}

Where N is the number of clouds with masses greater than M. 

\begin{figure*}%
    \centering
     {{\includegraphics[width= 8.5 cm ]
   { 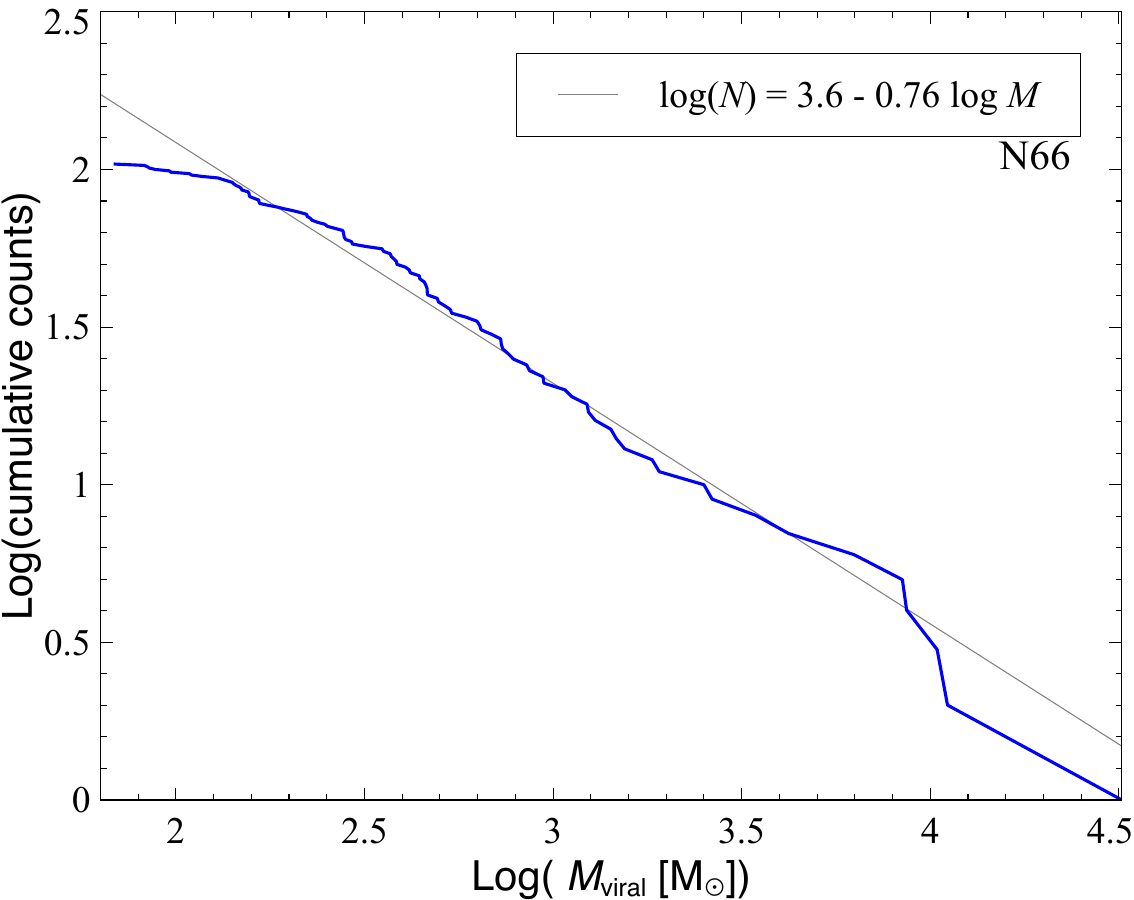} }}%
    \qquad
    {{\includegraphics[width= 8.5 cm]
    {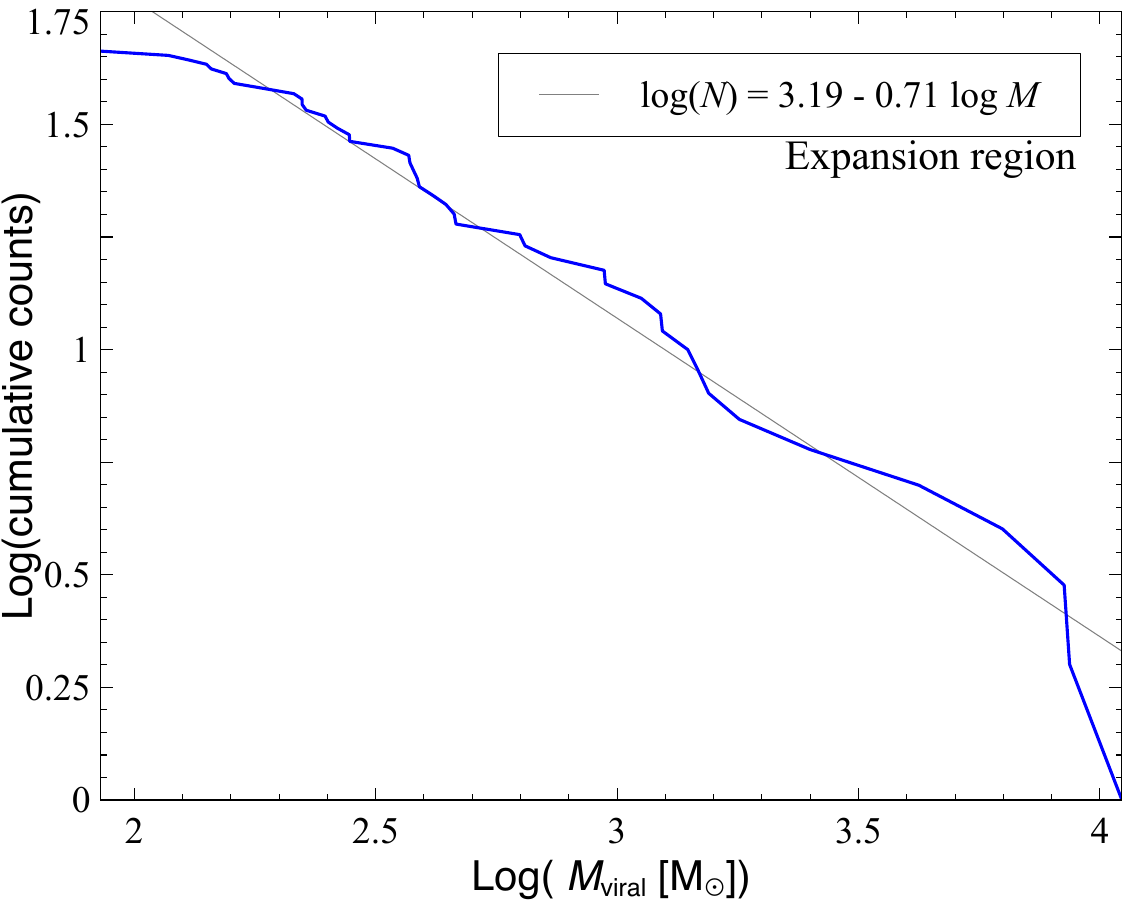} }}%
        \caption{The cumulative mass distribution of the \textsuperscript{12}CO trunks using the virial mass of: (a) the whole N66 with the best fit having a power law index of 0.76. (b) the expansion region indicated by an ellipse in Figure \ref{fig:IRAC+YSO+stars} with a power index of 0.71.}
        \label{fig:imf}
\end{figure*}

Using the clumps parameters obtained from the \textsc{Astrodendro} trunks, Figure \ref{fig:imf} represents cloud mass spectra in the cumulative form using the virial mass of the N66 clumps. The best fit of the whole region is given as,

%

\begin{equation}
\log (N (> M))= 3.6 - 0.76 \log (M) .
\end{equation}

With a power law index of 0.76, the $\alpha $ in N66 turns out to be 1.76. While the expansion region (indicated as an ellipse in Figure \ref{fig:IRAC+YSO+stars}) has a fit of, 

\begin{equation}
\log (N (> M))= 3.19 - 0.71 \log (M) .
\end{equation}

Resulting in the expansion region to have an $\alpha $ of 1.71.
The $\alpha $ in N66 as a whole and in the expansion region alone show similar values agreeing with the values of 1.76 \citep{Takekoshi17} and 1.7 \citep{Ohno23} obtained for the SMC. While \citet{Saldano23} has reported a steeper value of $\approx 2$ for the SMC.

\subsection{Kinematics} \label{ssec:Kinematics}

\begin{figure}
    \centering
	\includegraphics[width=17.5 cm]
     {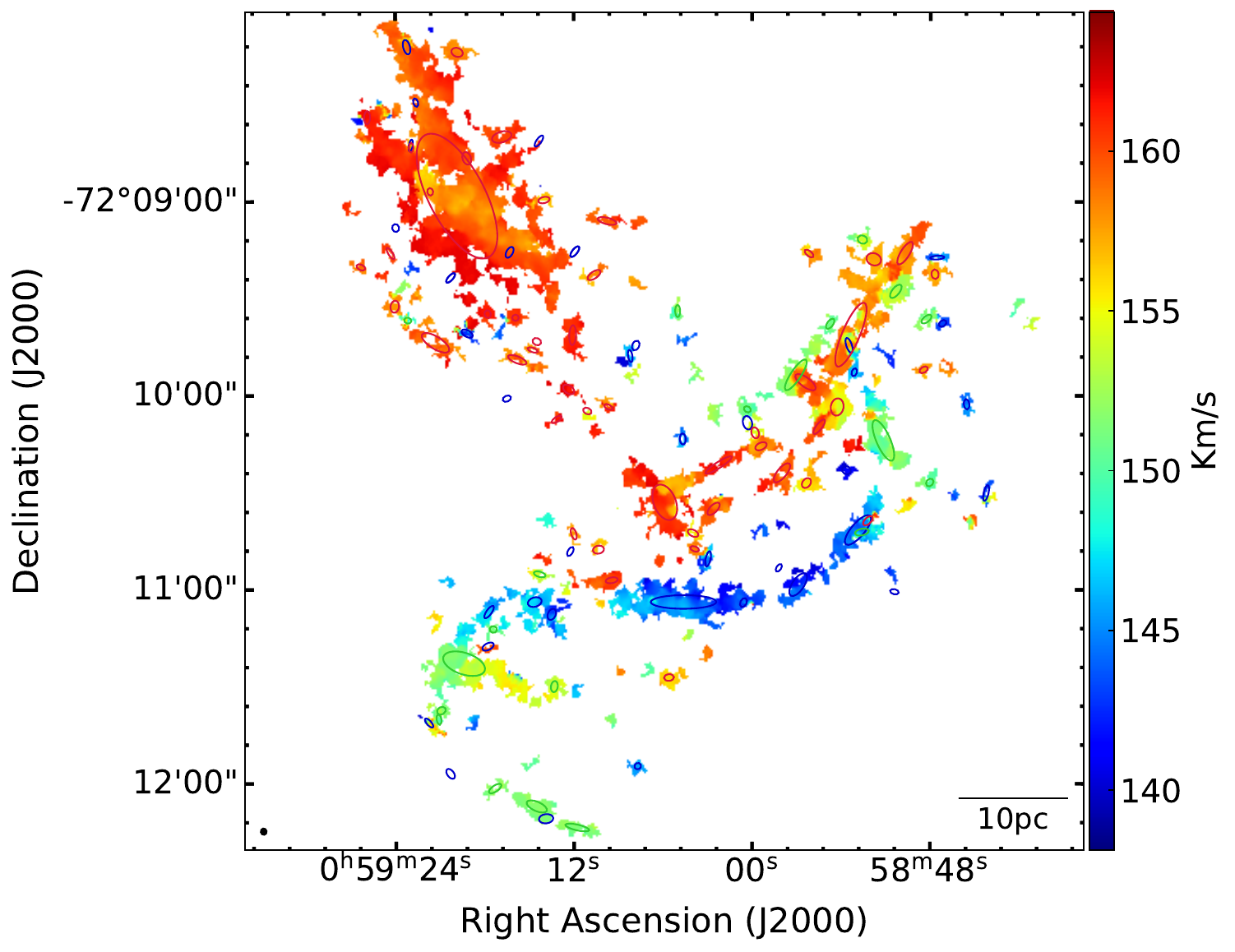}
    \caption{The \textsuperscript{12}CO(1–0) first-moment or intensity weighted velocity map of N66. The color scale represents the blue and red-shifted components for the region. The molecular clumps are shown as ellipses: blue ellipses for the velocities between 120-148 km s$^{-1}$, green ellipses for velocities between 148-154 km s$^{-1}$, while the red ellipses represent the 154-170 km s$^{-1}$ velocities.} 
    \label{fig:Moment1}

\end{figure}

To understand the kinematics of the \textsuperscript{12}CO(1-0) in N66, we inspect intensity-weighted velocity map (moment 1, Figure \ref{fig:Moment1}). There is a significant velocity distinction within the cloud movements. Some filaments are blue-shifted, while others show red-shifted velocities. Moreover, some other filaments show intermediate velocity components that appear in green in Figure \ref{fig:Moment1}. In the central region of N66, highlighted by an ellipse in Figure \ref{fig:IRAC+YSO+stars}, the blue-shifted filaments form a curved crescent shape. The red-shifted components seem to lie in the background, while the green components are observed along the edges. Together, these three features create a bubble-like structure.

\begin{figure}
    \centering
    \includegraphics[width=8.5 cm]
     {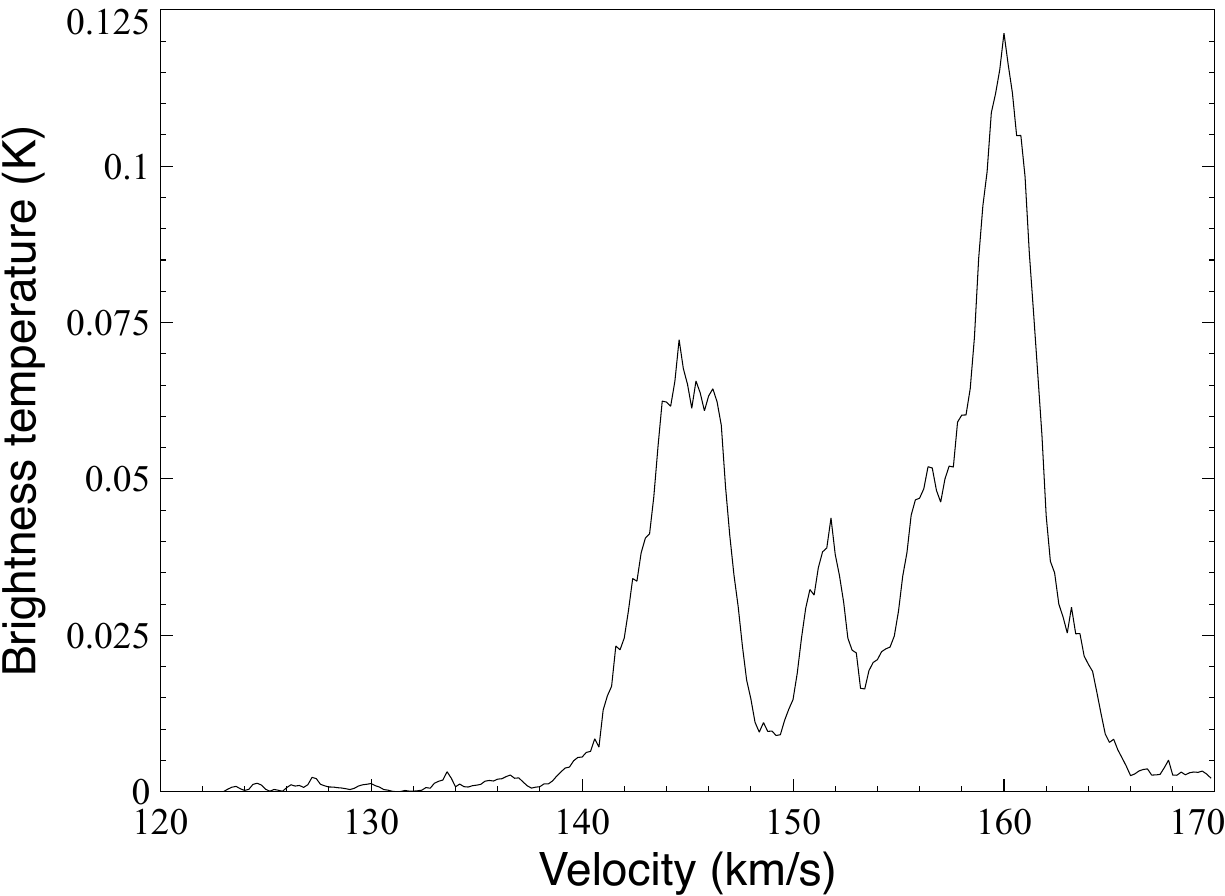}
    \caption{The average of all the \textsuperscript{12}CO(1–0)spectra (in $T^*_{\mathrm{A}}$) observed toward the central region, indicated by an ellipse in Figure \ref{fig:IRAC+YSO+stars}, of N66 with three main distinct peaks.   }
    \label{fig:spectra_peak}
\end{figure}

To better understand the kinematics of the \textsuperscript{12}CO(1-0) of this N66 central region, Figure \ref{fig:spectra_peak} shows the global averaged spectra of the elliptical region indicated in Figure \ref{fig:IRAC+YSO+stars}. The averaged spectra show three distinct velocity components. The strongest intensity peak is between the velocities 153-168 $\mathrm{km \, s^{-1}}$, with the highest point being at 160 $\mathrm{km \, s^{-1}}$. The second-largest peak at 145 $\mathrm{km \, s^{-1}}$, has an intensity approximately half that of the strongest peak and spans velocities from 138 $\mathrm{km \, s^{-1}}$ to 148 $\mathrm{km \, s^{-1}}$. The third and weakest peak is at 152 $\mathrm{km \, s^{-1}}$, covering a velocity range of 149 to 153 $\mathrm{km \, s^{-1}}$. The velocity center of the molecular clumps computed by \textsc{Astrodendro} align well with the red and blue shifted components appearing in the intensity-weighted velocity map (Figure \ref{fig:Moment1}). 


    

The presence of multiple peaks within a region is an indication that some interaction is occurring within the cloud (\citealt{Anderson15,Liu18,Baug19, Bhadari21}).

\subsection{Expansion Parameters} \label{sec:expansion}

We use the position-velocity (PV) diagram to better understand the kinematics of the region. To explore the possibility of an expanding shell in the region, a circular, ring-like, or U-shaped structure would be expected to appear in the PV diagram(e.g., \citealt{Moon98, Wei99,Tsuboi09,Butterfield18,Feddersen18}). 

\begin{figure}%
    \centering
    \gridline
   { \fig { 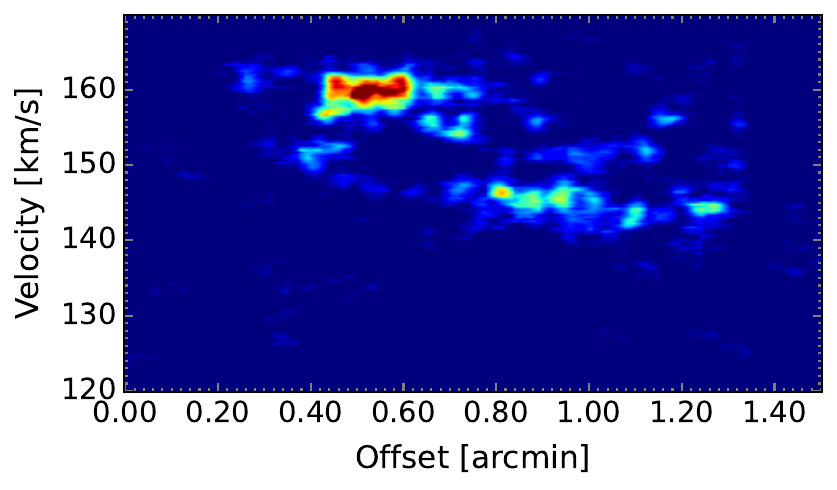} {0.45\textwidth} {(a) the cut along the minor axis of the ellipse, or the vertical cut.} 
     \fig { 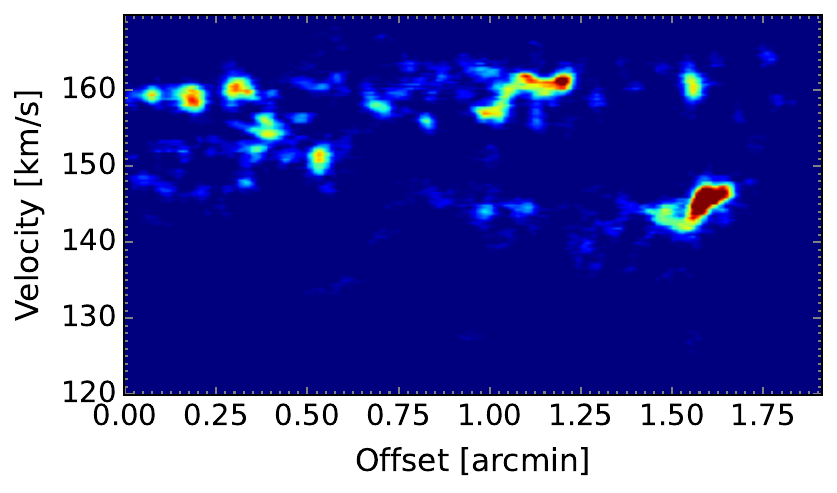} {0.45\textwidth} {  (b) cut along path a} } 
     
    \gridline
   { \fig  { 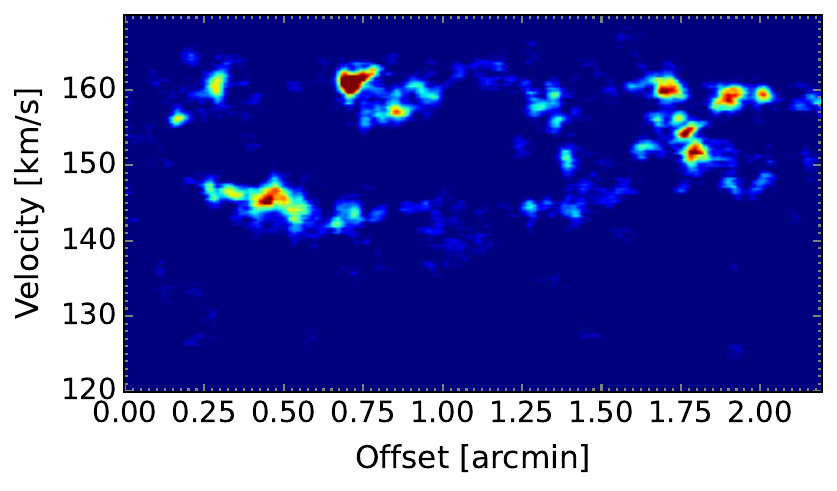} {0.45\textwidth} {(c) the cut along the major axis of the ellipse, or the horizontal cut.}   
    \fig  {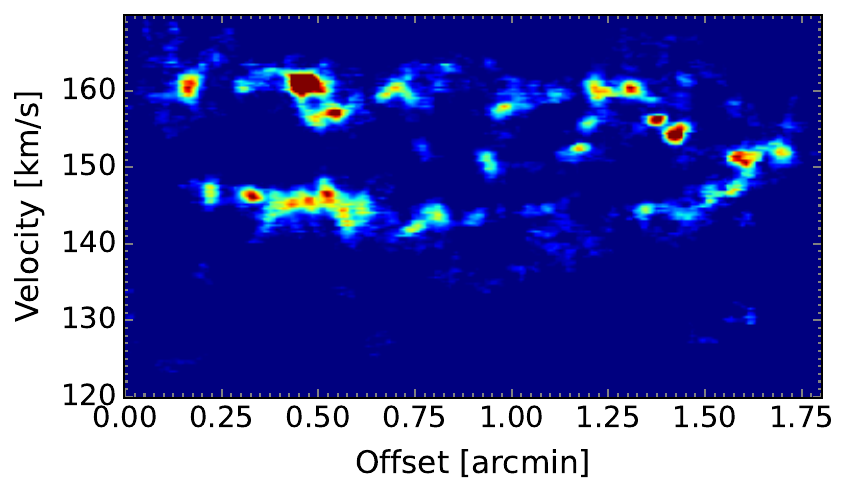} {0.45\textwidth} {(d) cut along path b}
    }
 \caption{The position-velocity maps along the cuts indicated in Figure \ref{fig:IRAC+YSO+stars}, with a circular ring-like feature appearing in of the maps.  }%
    \label{fig:PV}%
\end{figure}

In the PV diagrams, four spatial directions were chosen as indicated in Figure \ref{fig:IRAC+YSO+stars}. The line indicated with v represents the vertical cut which is the cut along the minor axis of the ellipse, and the h line represents the horizontal cut which is the major axis of the ellipse, while the a and b lines represent the tilted cuts of the bubble. The results of our \textsuperscript{12}CO(1–0) PV diagrams are shown in Figure \ref{fig:PV}. In all of these cuts, the PV diagrams showed a clear elliptical ring-like structure which indicates an expanding bubble. Using Figure \ref{fig:PV}, we are able to indicate the bubbles' approaching velocity to be $ v_{\mathrm{app}}\approx 140 \, $$\mathrm{km \, s^{-1}}$ (the lowest velocity in the ring-like structure) and a receding velocity of $v_{\mathrm{rec}}\approx 162 \,$ $\mathrm{km \, s^{-1}}$ (the highest velocity in the ring-like structure.)

 From the PV diagrams and using the geometry of an expanding shell \citep{Butterfield18} we can estimate the expanding shell properties. 
 
 An estimate of the velocity of the system can be obtained as,
 \begin{equation}
 v_{\mathrm{sys}}= \frac{v_{\mathrm{app}} + v_{\mathrm{rec}} } {2}  
 \end{equation}

 With the values of $v_{\mathrm{app}}$ and $v_{\mathrm{rec}}$ indicated earlier, the systemic velocity is estimated to be around  $v_{\mathrm{sys}}\approx  151 \,$ $\mathrm{km \, s^{-1}}$.
 
The expansion velocity is obtained by, 
 \begin{equation}
  v_{\mathrm{exp}}= \frac {v_{\mathrm{rec}}  -  v_{\mathrm{app}} }  {2}   
 \end{equation}
 
  As a result the expansion velocity of N66 bubble appears supersonic, $v_{\mathrm{exp}} \approx 11 \,$ $\mathrm{km \, s^{-1}}$. This value is similar to the expansion velocity ($v_{\mathrm{exp}} \approx [7-10] \,$ $\mathrm{km \, s^{-1}}$) estimated by \citet{Israel16} for the N66 bubble based on [C{\sc ii}] emission.

Similar to these geometric calculations, an estimate of the expanding bubble's radius can be deduced using the offset edges of the circular pattern in the PV diagrams. Using the vertical PV-cut (Figure \ref{fig:PV}a), an estimate of the radius of the shell is taken to be around $r_{\mathrm{v}} \approx 9.8 \pm 0.1 $ pc, while using the horizontal PV-cut (Figure \ref{fig:PV}c) the radius appears to be $r_{\mathrm{h}} \approx 12.9 \pm 0.5 $ pc instead. These values appear smaller than the [C{\sc ii}] linear radius of 25 pc found by \cite{Israel16}. In some clouds, the amount of molecular gas traced by [C{\sc ii}] surpasses that traced by \textsuperscript{12}CO \citep{Velusamy10}.
In the SMC, over 90$\%$ of the molecular material is not detected by CO \citep{Tokuda21}. This difference in the amount of molecular material being traced by [C{\sc ii}] compared with the \textsuperscript{12}CO can be the cause for the radius difference found.


With our value estimates and taking the assumption that the expanding velocity of this bubble is constant and does not vary with time (which is not a proper approximation of reality), we can estimate this expanding shell’s age to be around $t_{\mathrm{age}} \approx 1.2 \times 10^6$ years. While \citet{Israel16} interpreted a 3.5 Myr upper limit on the expansion age, with a 2 Myr estimate while taking the deceleration into account. These differences are a direct link to the radius estimate difference.


\section{Discussions } \label{sec:discu}

\subsection{Meaning Of The Size–Linewidth Relation }\label{sec:disc-scaling}

The comparison of the size-linewidth shown in Figure \ref{fig:Cloud_comparision} indicates that the N66 exhibits a fit with a power index similar to Orion, N55, and the SMC.

While the SMC-North clouds from \citet{Ohno23} exhibit lower velocity dispersions than those typically observed in the Milky Way, the N66 molecular cloud shows notably higher values. This trend of elevated linewidth is also observed in 30 Doradus \citep{Indebetouw13} a higher degree. Such an increase in velocity dispersion, exceeding the predictions of \citeauthor{Larson81}'s relation, can occur when clouds are in virial equilibrium but are influenced by external pressure \citep{Field11, Keto86}.

This external pressure and larger linewidth in N66 compared to other SMC and MW clouds can be due to: the energy injection and pressure coming from the cloud-cloud collision occurring in the N66 northern region \citep{Neelamkodan21}; the pressure from photoionization coming from the large number of high mass stars; the feedback processes and the pressure from the expansion bubble.

The analysis shows that N66 overall follows the size-linewidth relationship described by \citet{Larson81}, with a power index similar to that of Orion, N55, and the SMC. However, this agreement diverges when focusing solely on the expansion region of N66, highlighted as an ellipse in Figure \ref{fig:IRAC+YSO+stars}. In Figure \ref{fig:Cloud_comparision}, the power law fit of the clumps (trunks) velocity dispersion versus radius for the expansion region in N66 follows: 

$ \sigma_{\mathrm{v}}  = 0.90 \times  R^{ 0.75}$.


With a power-law index of 0.75, the size-linewidth relation in the expansion region aligns more closely with that of Galactic center clouds than with Orion or N55. This may be attributed to the expansion within the region, which perturbs the cloud and exposes it to increased external pressure.

{
The Galactic center, with its extreme environment characterized by high-energy radiation, tidal forces, and significant external pressures, creates conditions that lead to steeper size-linewidth relations in molecular clouds. This is a result of the complex interactions that drive enhanced turbulence and velocity gradients within the clouds. These turbulent conditions are often stronger than those in typical star-forming regions, such as those observed in the Orion nebula. This trend can also be visible in other high-pressure environments like cloud-cloud interactions and feedback processes that significantly affect the size-linewidth relationships \citep{Green24}.

The similarity of the size-linewidth relation of N66's expansion region to that of clouds in the Galactic center and 30 Doradus suggests that the dynamics within N66 are strongly influenced by these external pressures and feedback processes, likely leading to more intense turbulence compared to typical star-forming regions.

 An alternative possibility for the larger linewidth, especially in the expansion region, is that the clumps there undergo gravitational collapse and deviate from virial equilibrium \citep{Ballesteros-Paredes18}}.

\subsection{The X$_{CO}$ Factor} \label{sec:disx_X}

The $X_{\mathrm{CO}}$ value for the MW is $X_{\mathrm{CO}}^{\mathrm{MW}}= 2 \times 10 ^{20}\mathrm{ \, cm^{-2} (K\,km \,s^{-1})^{-1} }$ 
\citep{Dame01,Bolatto13}. For the SMC-North, \citeauthor{Ohno23}'s calculation for the $X_{\mathrm{CO}}$ factor, under the assumption of virial equilibrium, was determined to be  $1.3  \times 10 ^{21} \mathrm{ \, cm^{-2} (K \, km \,s^{-1})^{-1}} $ for the clumps, indicating that the SMC region has a greater conversion factor than that of the MW. The N66 clouds with $L_{\mathrm{CO}} > 4 \, \mathrm{K \, km \,s^{-1} pc^2}$ has a slightly higher value of $X_{\mathrm{CO}}$ than that for the SMC-North($ 3.26 \times 10 ^{21}\mathrm{ \, cm^{-2} (K \, km \,s^{-1})^{-1} }$), while the $X_{\mathrm{CO}}$ for the $L_{\mathrm{CO}} < 4 \, \mathrm{K \, km \,s^{-1} pc^2}$ contributes to an even higher $X_{\mathrm{CO}}= 7.6 \times 10 ^{21}\mathrm{ \, cm^{-2} (K \, km \,s^{-1})^{-1} }$. Several studies have been done on the $X_{\mathrm{CO}}$ in the SMC.
The large-scale SMC surveys by \citet{Rubio91} suggest an $X_{\mathrm{CO}}= 6 \times 10 ^{21}\mathrm{ \, cm^{-2} (K \, km \,s^{-1})^{-1} }$, while the survey by \citet{Mizuno01} suggests an  $X_{\mathrm{CO}}= 2.5 \times 10 ^{21}\mathrm{ \, cm^{-2} (K \, km \,s^{-1})^{-1} }$, in which our results are also similar to.

\citet{Bolatto13} have suggested that the value of $X_{\mathrm{CO}}$ is significantly influenced by the size of the cloud. It is expected to fluctuate significantly on smaller scales while averaging out on larger scales. These smaller-scale clouds would exhibit an elevated level of $X_{\mathrm{CO}}$. This trend is also observed in our N66 data, where smaller clouds and those with lower luminosities exhibit a notable increase in $X_{\mathrm{CO}}$ compared to the larger, more luminous regions of the cloud. This behavior aligns with previous findings by \citet{Bolatto13}.

Dust-based calculations of $X_{\mathrm{CO}}$ imply an increase in the factor as metallicity decreases. However, $X_{\mathrm{CO}}$ calculations using virial mass on small scales contradict the results obtained from dust-based observations \citep{Bolatto13}. Moreover, \citet{Bolatto13} pointed out that the virial mass based $X_{\mathrm{CO}}$ derivation probably overestimates the value in the weaker CO regions, indicating that at high spatial resolution, virial mass measurements would approximate that of the Galactic $X_{\mathrm{CO}}$. With our higher resolution study in N66 using the CO data, our $X_{\mathrm{CO}}$ factor appears to have a higher value than that of the Galactic. Note that the link of $X_{\mathrm{CO}}$ and metallicity is dependent on how the relative sizes of the [C{\sc  ii}] and CO-emitting regions vary with lower heavy element and dust abundance \citep{Bolatto13}.

It is important to recognize that the $X_{\mathrm{CO}}$ factor, which influences the mass estimates of galaxies, is not solely dependent on metallicity but can also be significantly affected by interstellar pressure, feedback mechanisms, and variations in density. The cloud properties estimates are sensitive to the cloud size, constraints, and quality of the data, making $\alpha _{\mathrm{CO}}$ and $ X _{\mathrm{CO}}$ for N66 highly uncertain.




\subsection{What Drives The Expansion Bubble} \label{sec:drivesit}

\begin{figure}%
    \centering
    \includegraphics[width = 17.5 cm]
    {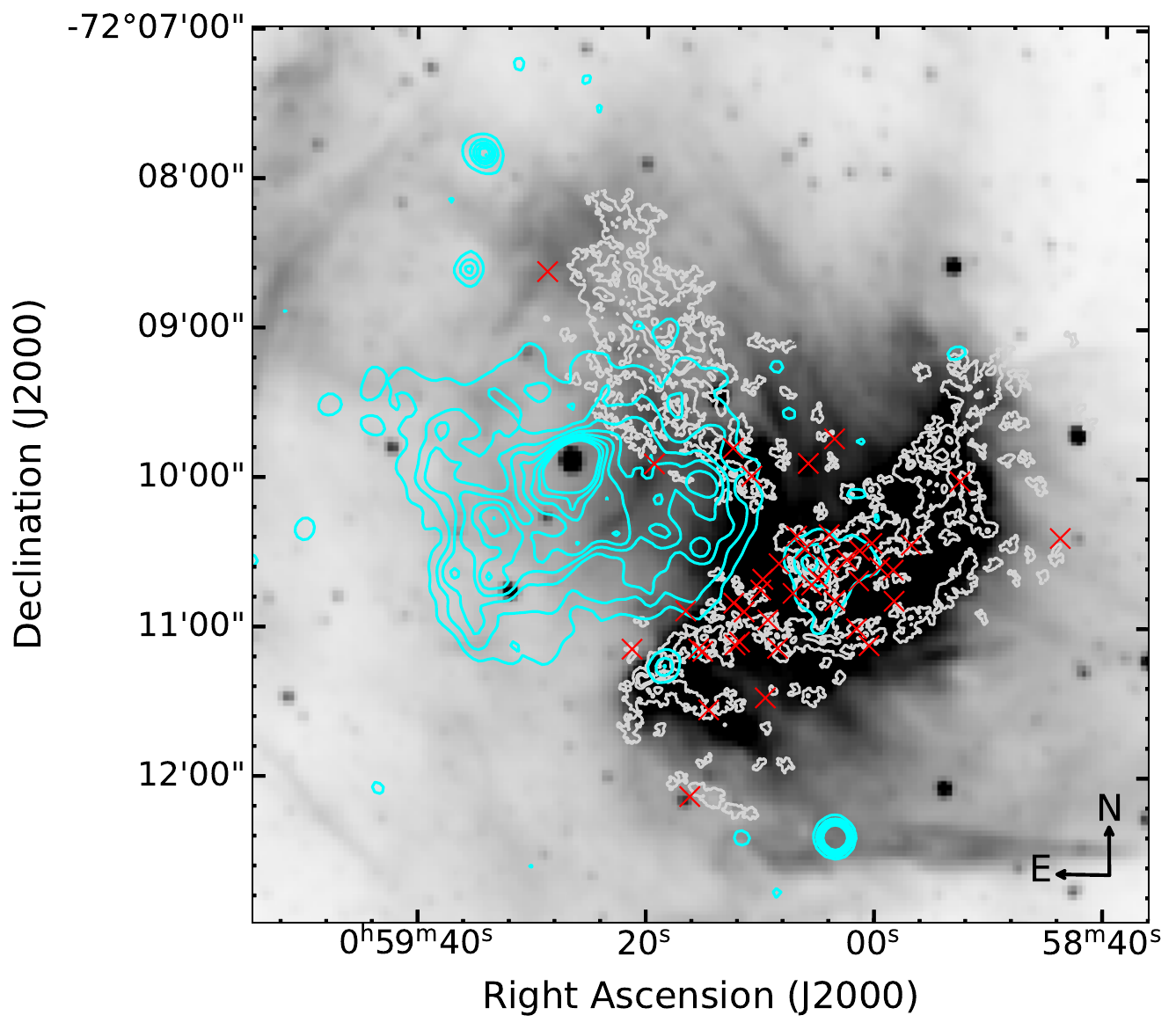}
    \caption{The H$_{\mathrm{\alpha}}$ map \citep{Smith98} of N66 with the \textsuperscript{12}CO(1–0) gray contours at level 0, 20, 40, and 80 K km s$^{-1}$ overlaid. The OB stars \citep{Dufton19} are indicated in red crosses (x). The X-ray contours of the SNR B0057-724 \citep{Naze02} are shown in cyan at an interval of $3\times 10^{-8}$ photons cm$^{-2}$ s$^{-1}$. }%
    \label{fig:Ha}%
\end{figure}

The N66 nebula has an interesting composition with a variety of stellar populations embedded within it. When investigating the driving factor of the expansion bubble within N66, the first thing that comes to mind is whether there is a supernova remnant (SNR) that might have triggered it. The N66 hosts 3 main SNRs: SNR B0056-724, SNR B0056-725, and SNR B0057-724 (SNR 0057-7226) \citep{Ye91, Reid06}, where the first two SNRs are located further away from the field studied. The SNR B0057-724 (also referred to as IKT 18), shown in Figure \ref{fig:Ha}, is the nearest to the field. High-velocity features have been identified across its location, suggesting an expanding shell associated with it \citep{Danforth03}. 
 
 The X-ray contours of SNR B0057-724 appear to be within the molecular cloud (Figure \ref{fig:Ha}). 
 While the SNR influences the molecular cloud to some extent, it is unlikely to be the primary driver of the expansion observed in the N66 bubble. Instead, N66 seems to have an expanding motion independent of the SNR shocks. \citet{Gouliermis08} proposed that the expansion of the SNR in the outer region of N66 may have triggered the star formation away from the central region, in contrast the recent burst of star formation in the central region is likely dominated by photoionization feedback from the OB stars located there. This distinction suggests that different mechanisms are driving star formation in the central and peripheral regions of N66.

 Figure \ref{fig:Ha} shows that most of the massive stars in N66 are surrounded by an elevated amount of H${\mathrm{\alpha}}$ and are actually found in the region of expansion. This raises the question of whether there is a correlation between the massive stars and the expansion region they occupy. One possibility is that the massive stars and their radiation are triggering a new generation of star formation in the surrounding gas. Alternatively, the expansion bubble could be the result of these stars, with their radiation and feedback playing a key role in shaping the surrounding gas. 


\citet{Gouliermis14} indicates that NGC346 has at least two distinct cluster populations, one that is extended and has formed hierarchically, and the other being a centrally condensed young cluster. The age of those stellar populations does increase with radial distance from the central region \citep{Dufton19,Sabbi08}, with more massive stars being at the center. The massive stars in the inner region of N66 have ages less than 2 Myr, with an uncertainty of about 1-2 Myr. This places some of the stars in the central region close to the zero-age main sequence (ZAMS) \citep{Dufton19}. Given that the young massive stars within the expansion bubble have ages similar to that of the bubble itself, it is possible that these stars and the recent burst of star formation in the central region are linked to the expansion bubble. This correlation suggests that the feedback from star formation may play a role in driving the expansion.

The relationship between the expansion bubble, the massive stars, and the radiation they emit can be viewed as a feedback loop. Massive stars form and emit intense radiation, which drives the formation of the expansion bubble. This bubble, in turn, triggers further star formation in the region through positive feedback, sustaining the cycle of star formation and feedback-driven gas dynamics.

\citet{Lopez14} have investigated the effects of stellar feedback on the larger scale  N66 H{\sc ii} region, finding that the direct radiation pressure from the massive stars in the region does not play a significant role currently in the dynamics, with the warm gas pressure, $P$\textsubscript{H{\sc ii}},
being dominant. It is important to note that \citet{Lopez14} has investigated the effect of feedback on N66 at a further distance than the expansion bubble studied in this paper.  

To look deeper into the effect that the direct radiation pressure ($P_{\mathrm{dir}}$) from the massive stars has at a closer scale, we use \citet{Lopez14} volume-averaged direct radiation pressure:

\begin{equation}
P_{\mathrm{dir}} = \frac { 3 L_{\mathrm{bol}} }  {4\pi R^2c} 
\end{equation} 

Where $R$ is the region radius, and $ L_{\mathrm{bol}}$ is the bolometric luminosity. Using the same estimate from \citet{Lopez14}, we get the IMF  bolometric luminosity as:
 
\begin{equation}
L_{\mathrm{bol,IMF}} \approx 138 L_{\mathrm{H\alpha}}
\end{equation} 

\begin{figure}
    \centering
    \includegraphics[width=8.5 cm]
     {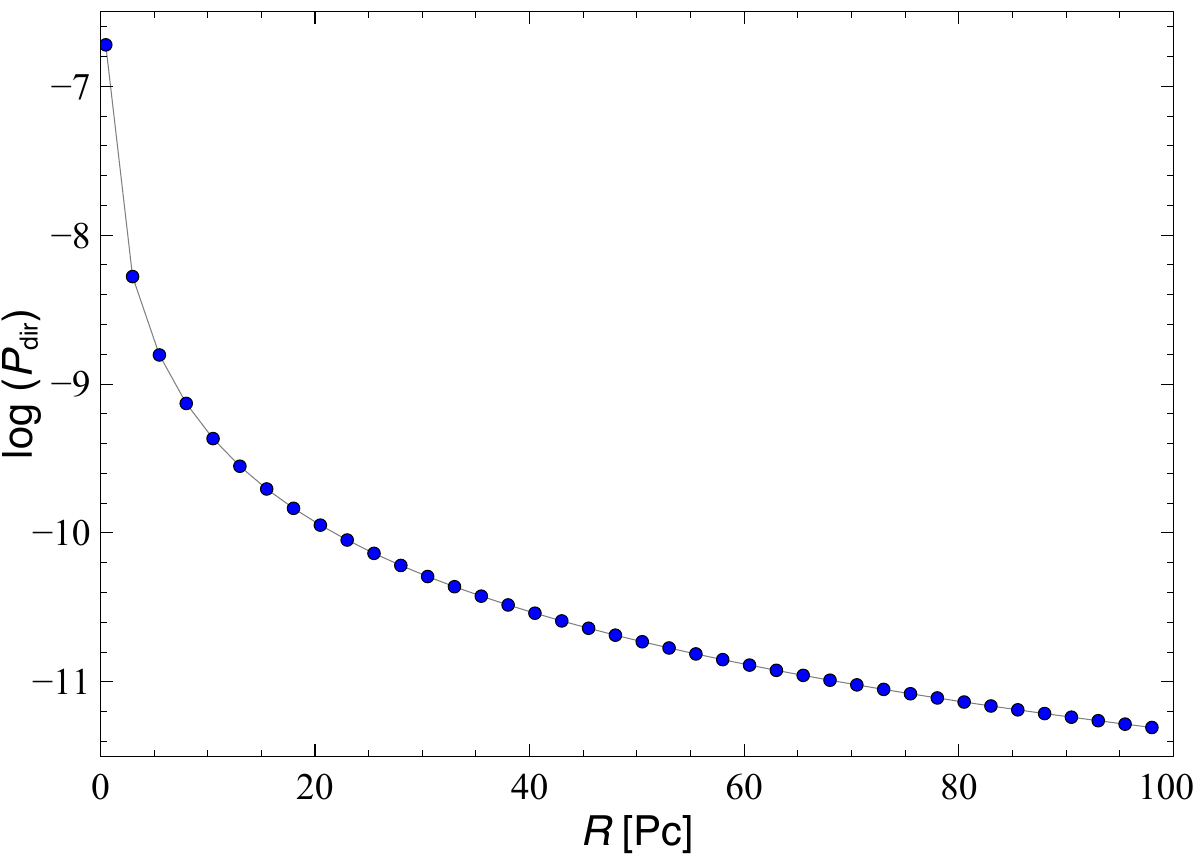}
    \caption{ The way the direct pressure from the stars and the radius from the center of N66 affect one another. It is apparent that the $P_{\mathrm{dir}}$ has a major effect at the closer distance, and its effect drops drastically after some distance.  }
    \label{fig:P_dir}
\end{figure}

 With a total $\mathrm{H\alpha}$ luminosity in N66 region being $L(\mathrm{H\alpha})=4.1 \times 10^{38} \mathrm{ erg \, s^{-1} }$\citep{Geist22}. In Figure \ref{fig:P_dir}, we show the variation of $P_{\mathrm{dir}}$ with radius in the N66. The $P_{\mathrm{dir}}$ seems to play a more significant role in the first 20 pc. We find that the $P_{\mathrm{dir}}$ at the expansion bubble edge is $P_{\mathrm{dir}}(bubble) = [ 4.94 - 2.85 ]  \times 10^{-10}$ dyn cm$^{-2}$.

To better understand the dominant factor driving the expansion, \citet{Lopez14} provides an approximate calculation of the characteristic radius ($r_{\mathrm{ch}}$), which marks the transition from radiation pressure dominance to gas pressure dominance within the region. This $r_{\mathrm{ch}}$ reduces to, 

\begin{equation}
r_{\mathrm{ch}} = 0.072 \left( \frac {S} {10^{-49}\mathrm{s}} \right) \mathrm{ pc}
\end{equation}

Here S is the ionizing photon luminosity, which is  $S \approx 3 \times 10^{50} \, \mathrm{s}^{-1}$ \citep{Geist22} in the region. This leads to an estimate of $r_{\mathrm{ch}} \approx 2.15$ pc. The expanding shell, currently at around 10 pc, indicates that radiation pressure likely played a dominant role in the past. However, based on current estimates, the expansion is now primarily driven by gas pressure. It is important to note that these estimates assume that the radiation pressure from the stars continues to originate from the expansion center. In reality, the stars are closer to the expansion shell edge than that. The N66 expansion shell can be initially driven by the direct radiation pressure from the stars, then as the shell expanded and led to compression of gas on its edges, more massive stars began forming there, leading to more direct radiation pressure closer to the edge, and fueling the expansion further.

\section{Summary and Conclusions} \label{sec:sum}

We report the \textsuperscript{12}CO(1–0) properties of the molecular cloud in the star-forming region N66 in the SMC using ALMA observations. The analysis and results are summarized as follows: 

\begin{enumerate}
    \item  N66's CO cloud has a clumpy, filamentary structure on sub-parsec scales. Using \textsc{Astrodendro}, a python package that identifies unique isosurfaces from a region of emission in a three-dimensional data cube, the cloud's structure has been divided into clumps (trunks) and cores (leaves) with a total of 165 substructures: 104 of which are clumps (trunks), while 61 of which were molecular cores (leaves). The clumps and cores were used separately to understand their physical properties like radius, linewidth, and mass. 

    \item The size-linewidth relation for the region as a whole has a power index that follows the \citet{Larson81} scaling relationship well. The power law fit for the molecular clumps (trunks) alone results in: $ \sigma_{\mathrm{v}}  = 0.85 \times  R^{ 0.54}$, which aligns well with the 0.5 power index scaling relationship. However, the cloud as a whole has a higher linewidth than that of the SMC and even the MW, which might be due to external pressure. When investigating the size-linewidth of the central N66 region alone, the clumps have a power fit of $ \sigma_{\mathrm{v}}  = 0.90 \times  R^{ 0.75}$. This greater power indicates that there is a different factor that affects this region alone.

    \item The CO virial mass-luminosity scaling relation for N66 trunks follows a $M\textsubscript{vir}=181  \, L \, ^{0.67} $ power fit. With that, an estimate for the $X_{\mathrm{CO}}$ factor on the larger scale was found to be  $ X_{\mathrm{CO}}  =  3.26 \times 10^{21} \mathrm{ cm^{-2} (K \, km \,s^{-1})^{-1}} $ which is comparable to that obtained for the SMC-North.

    \item The observation of the central N66 region indicated that the average of all the \textsuperscript{12}CO(1–0) spectra has three main distinct peaks with velocity ranges: [138-148 $\mathrm{km \, s^{-1}}$] peaking at 145 $\mathrm{km \, s^{-1}}$, [149 - 153 $\mathrm{km \, s^{-1}}$] peaking at 152 $\mathrm{km \, s^{-1}}$, and [ 153 - 168 $\mathrm{km \, s^{-1}}$ ] peaking at 160 $\mathrm{km \, s^{-1}}$. A circular structure appears in the PV diagram of all the cuts taken at this central N66 region. This characteristic is an indication that this specific region has an expanding shell. The bubble appears supersonic with an expansion velocity around $v_{\mathrm{exp}} \approx 11$ $\mathrm{km \, s^{-1}}$,  with an overall systemic velocity of $v_{\mathrm{sys}}\approx $ 151 $\mathrm{km \, s^{-1}}$. The expansion region radius estimate is $r_{\mathrm{v}} \approx 9.8- 12.9 \pm 0.5$ pc. The expanding shell's age is estimated to be around $t_{\mathrm{age}} \approx 1.2 \times 10^6$ years.  
    
    \item The expanding bubble's driving mechanism was investigated. It was determined that at an earlier stage, when the bubble was smaller, direct radiation pressure could have dominated the region and thus could have triggered this expansion. With a radius estimate of $r_{\mathrm{ch}} \approx 2.15$ pc at which the source transitions from being radiation pressure dominant to being gas pressure dominant. At this stage, the expansion bubble is not dominant in radiation pressure.

\end{enumerate}

\begin{acknowledgments}

This paper makes use of the following ALMA data: ADS/ JAO.ALMA $\#$ 2015.1.01296.S. ALMA is a partnership of the ESO, NSF, NINS, NRC, NSC, and ASIAA.The Joint ALMA Observatory is operated by the ESO, AUI/NRAO, and NAOJ. This research has been supported by United Arab Emirates University (UAEU), under UPAR grant Codes G00003479 and G00004651. Tokuda acknowledges that this work was supported by a NAOJ ALMA Scientific Research grant No. 2022-22B, and Grants-in-Aid for Scientific Research (KAKENHI) from the Japan Society for the Promotion of Science (JSPS; grant Nos. JP20H05645, JP21H00049, and JP21K13962).
\end{acknowledgments}

%



\appendix

\begin{longtable}{llllllllllll}
\caption{\textsuperscript{12}CO(1–0) Clump (Trunks) Properties. \\
}\\

\hline
ID  &  RA      & Dec.     & $v_{\mathrm{cen}}$  & $\sigma_{\mathrm{maj}}$   & $\sigma_{\mathrm{min}}$  & Angle \textsuperscript{1} & $v_{\mathrm{rms}} $ & Flux \textsuperscript{2}  & Radius  & $M_{\mathrm{vir}}$  & $L_{\mathrm{CO}}$  \\
   & (deg)     & (deg)     & (km s$^{-1}$)  & (arcsec) &  (arcsec) &  (deg) & (km s$^{-1}$) & ( K ) & ( pc ) &  ($\mathrm{M_{\odot}}$) &  (K km s$^{-1}$ pc$^2$)   \\ \hline
\endfirsthead 
\caption[]{(Clump properties continued)}\\

\hline
ID  &  RA      & Dec.     & $v_{\mathrm{cen}}$  & $\sigma_{\mathrm{major}}$  & $\sigma_{\mathrm{minor}}$  & Angle \textsuperscript{1}  & $v_{\mathrm{rms}} $ & Flux \textsuperscript{2}  & Radius  & $M_{\mathrm{vir}}$  & $L_{\mathrm{CO}}$\\
   & (deg)     & (deg)     & (km s$^{-1}$)  & (arcsec) &  (arcsec) &  (deg) & (km s$^{-1}$) & ( K ) & ( pc ) &  ($\mathrm{M_{\odot}}$) &  (K km s$^{-1}$ pc$^2$)  \\ \hline
\endhead

\hline 
\multicolumn{12}{l}{\textsuperscript{1} The orientation angle of the major axis.} \\
\multicolumn{12}{l}{\textsuperscript{2} The integrated flux of a structure.}

\endlastfoot

        1 & 14.7999 & -72.1544 & 124.2 & 1.98 & 0.81 & 51.9 & 0.79 & 11.3 & 0.70 & 455 & 0.99  \\ 
        2 & 14.8348 & -72.1566 & 124.4 & 1.87 & 0.74 & 47.7 & 0.35 & 11.2 & 0.65 & 82 & 0.98  \\ 
        3 & 14.8244 & -72.1883 & 126.3 & 1.91 & 1.03 & -152.5 & 0.74 & 11.5 & 0.78 & 443 & 1.01  \\ 
        4 & 14.8349 & -72.1992 & 126.0 & 1.76 & 1.03 & 127.3 & 0.44 & 24.9 & 0.74 & 150 & 2.18  \\ 
        5 & 14.8012 & -72.1801 & 126.7 & 1.58 & 0.78 & 60.4 & 0.49 & 17.2 & 0.61 & 155 & 1.50  \\ 
        6 & 14.8190 & -72.1670 & 128.3 & 1.35 & 0.91 & -156.7 & 1.04 & 18.5 & 0.61 & 687 & 1.62 \\ 
        7 & 14.8182 & -72.1544 & 128.3 & 1.85 & 0.99 & 60.8 & 0.61 & 13.2 & 0.75 & 292 & 1.15  \\ 
        8 & 14.7829 & -72.1624 & 128.6 & 1.59 & 1.06 & 64.6 & 0.79 & 15.2 & 0.72 & 464 & 1.33  \\ 
        9 & 14.7103 & -72.1836 & 130.4 & 1.36 & 0.80 & 166.8 & 0.61 & 35.0 & 0.58 & 223 & 3.06  \\ 
        10 & 14.8445 & -72.1415 & 130.4 & 1.33 & 0.75 & 107.3 & 0.41 & 12.0 & 0.55 & 96 & 1.05 \\ 
        11 & 14.8100 & -72.1449 & 131.7 & 2.04 & 0.77 & 54.8 & 0.61 & 29.4 & 0.69 & 265 & 2.57  \\ 
        12 & 14.8081 & -72.2031 & 132.9 & 2.22 & 1.43 & -173.8 & 0.59 & 34.6 & 0.99 & 355 & 3.03  \\ 
        13 & 14.8501 & -72.1523 & 134.3 & 1.24 & 1.08 & 104.5 & 0.86 & 13.7 & 0.64 & 493 & 1.20  \\ 
        14 & 14.7515 & -72.1691 & 134.5 & 2.13 & 1.41 & 104.4 & 0.53 & 10.4 & 0.96 & 279 & 0.91  \\ 
        15 & 14.8459 & -72.1452 & 134.7 & 1.83 & 0.60 & 80.6 & 0.57 & 10.4 & 0.58 & 194 & 0.91  \\ 
        16 & 14.7526 & -72.1845 & 135.6 & 1.50 & 0.98 & 64.5 & 0.52 & 10.2 & 0.67 & 186 & 0.89  \\ 
        17 & 14.7427 & -72.1815 & 136.5 & 1.34 & 0.74 & 55.8 & 0.50 & 17.8 & 0.55 & 141 & 1.56 \\ 
        18 & 14.7373 & -72.1830 & 142.1 & 4.13 & 1.81 & 56.0 & 2.35 & 246.6 & 1.51 & 8656 & 21.57 \\ 
        19 & 14.8410 & -72.1949 & 139.3 & 1.77 & 0.71 & 131.7 & 0.47 & 19.0 & 0.62 & 142 & 1.66 \\ 
        20 & 14.7695 & -72.1845 & 144.7 & 10.14 & 2.11 & -179.9 & 1.78 & 3498.4 & 2.56 & 8434 & 305.97\\ 
        21 & 14.8470 & -72.1368 & 140.5 & 2.30 & 1.05 & 104.3 & 0.74 & 18.2 & 0.86 & 497 & 1.59 \\ 
        22 & 14.7237 & -72.1732 & 140.7 & 1.46 & 1.25 & 153.4 & 0.52 & 12.0 & 0.75 & 214 & 1.05  \\ 
        23 & 14.7205 & -72.1783 & 144.8 & 5.82 & 2.04 & 49.8 & 1.46 & 597.9 & 1.91 & 4216 & 52.29  \\ 
        24 & 14.8065 & -72.1855 & 143.4 & 1.83 & 1.21 & 65.4 & 0.94 & 119.1 & 0.82 & 762 & 10.42  \\ 
        25 & 14.8302 & -72.1614 & 142.8 & 1.91 & 1.01 & 150.6 & 0.61 & 16.1 & 0.77 & 294 & 1.41  \\ 
        26 & 14.6846 & -72.1751 & 143.4 & 2.48 & 0.76 & 77.3 & 0.44 & 20.5 & 0.76 & 156 & 1.80  \\ 
        27 & 14.6968 & -72.1605 & 143.1 & 1.29 & 0.91 & -143.9 & 0.33 & 12.1 & 0.60 & 69 & 1.05  \\ 
        28 & 14.6983 & -72.1549 & 144.8 & 2.18 & 0.59 & -176.3 & 0.95 & 16.2 & 0.63 & 588 & 1.42  \\ 
        29 & 14.8241 & -72.1853 & 146.5 & 2.28 & 0.81 & 54.6 & 1.04 & 191.2 & 0.75 & 851 & 16.73  \\ 
        30 & 14.8113 & -72.1845 & 146.9 & 2.19 & 1.45 & -163.0 & 1.34 & 466.3 & 0.99 & 1831 & 40.78 \\ 
        31 & 14.6900 & -72.1675 & 144.8 & 1.62 & 0.84 & 96.8 & 0.69 & 16.6 & 0.65 & 318 & 1.45  \\ 
        32 & 14.7697 & -72.1705 & 145.2 & 1.70 & 0.93 & 93.5 & 0.80 & 9.7 & 0.70 & 460 & 0.85  \\ 
        33 & 14.7844 & -72.1633 & 145.8 & 1.92 & 0.67 & 99.1 & 0.77 & 10.1 & 0.63 & 386 & 0.89  \\ 
        34 & 14.7823 & -72.1986 & 145.6 & 1.04 & 0.98 & -153.8 & 0.53 & 43.4 & 0.56 & 166 & 3.80  \\ 
        35 & 14.7216 & -72.1647 & 146.3 & 1.27 & 0.81 & 70.8 & 0.52 & 43.8 & 0.56 & 157 & 3.83  \\ 
        36 & 14.7230 & -72.1624 & 147.5 & 2.41 & 0.87 & 108.7 & 0.87 & 135.9 & 0.80 & 629 & 11.88  \\ 
        37 & 14.7626 & -72.1808 & 146.2 & 2.41 & 0.75 & 78.0 & 0.57 & 10.3 & 0.75 & 248 & 0.90  \\ 
        38 & 14.8311 & -72.1898 & 152.2 & 6.74 & 3.33 & 161.6 & 1.95 & 3710.0 & 2.62 & 10427 & 324.48\\ 
        39 & 14.7134 & -72.1706 & 150.7 & 6.81 & 2.14 & 113.7 & 1.69 & 996.4 & 2.11 & 6281 & 87.14  \\ 
        40 & 14.7516 & -72.1679 & 150.5 & 1.15 & 0.90 & 147.3 & 1.12 & 206.5 & 0.56 & 729 & 18.06 \\ 
        41 & 14.8468 & -72.1603 & 148.8 & 1.13 & 0.92 & 164.5 & 0.69 & 22.9 & 0.57 & 278 & 2.00 \\ 
        42 & 14.7004 & -72.1742 & 150.3 & 1.33 & 1.04 & 51.8 & 0.77 & 63.5 & 0.65 & 406 & 5.55 \\ 
        43 & 14.7380 & -72.1650 & 152.0 & 5.67 & 1.54 & 56.4 & 0.74 & 346.0 & 1.64 & 940 & 30.26 \\ 
        44 & 14.7993 & -72.2038 & 151.6 & 3.77 & 0.88 & 167.3 & 0.71 & 158.1 & 1.01 & 534 & 13.83 \\ 
        45 & 14.8382 & -72.1946 & 150.2 & 1.63 & 0.72 & 99.2 & 0.47 & 37.5 & 0.60 & 137 & 3.28 \\ 
        46 & 14.8107 & -72.2020 & 151.6 & 3.39 & 1.44 & 156.3 & 0.57 & 455.0 & 1.22 & 419 & 39.79 \\ 
        47 & 14.7194 & -72.1785 & 151.3 & 2.16 & 1.02 & -167.9 & 0.72 & 31.3 & 0.82 & 442 & 2.74  \\ 
        48 & 14.7283 & -72.1606 & 150.6 & 1.81 & 0.91 & 53.4 & 0.40 & 21.6 & 0.71 & 118 & 1.89  \\ 
        49 & 14.7013 & -72.1602 & 151.7 & 1.84 & 0.91 & -141.3 & 0.61 & 21.8 & 0.72 & 282 & 1.91  \\ 
        50 & 14.8375 & -72.1938 & 152.1 & 1.46 & 1.14 & -148.3 & 0.79 & 47.3 & 0.71 & 465 & 4.14  \\ 
        51 & 14.8229 & -72.1868 & 151.0 & 1.13 & 1.05 & 163.7 & 0.38 & 13.7 & 0.60 & 88 & 1.19  \\ 
        52 & 14.8058 & -72.1917 & 153.7 & 1.73 & 1.05 & 81.4 & 0.97 & 110.0 & 0.75 & 727 & 9.62  \\ 
        53 & 14.7711 & -72.1595 & 152.0 & 1.86 & 0.77 & 90.9 & 0.57 & 14.6 & 0.66 & 227 & 1.28  \\ 
        54 & 14.7099 & -72.1578 & 153.1 & 2.50 & 1.14 & 51.2 & 0.94 & 348.1 & 0.94 & 866 & 30.44 \\ 
        55 & 14.7193 & -72.1533 & 153.3 & 1.57 & 1.23 & 156.9 & 1.27 & 55.9 & 0.77 & 1294 & 4.89 \\ 
        56 & 14.8224 & -72.2005 & 151.6 & 2.23 & 0.94 & -145.9 & 0.32 & 25.4 & 0.80 & 84 & 2.22 \\ 
        57 & 14.7264 & -72.1677 & 154.9 & 2.71 & 1.98 & 80.1 & 1.05 & 795.5 & 1.28 & 1471 & 69.57\\ 
        58 & 14.8099 & -72.1821 & 153.8 & 1.85 & 0.82 & 164.4 & 0.85 & 15.2 & 0.68 & 515 & 1.33 \\ 
        59 & 14.7494 & -72.1699 & 154.3 & 1.75 & 1.15 & 106.2 & 0.44 & 10.8 & 0.78 & 161 & 0.95 \\ 
        60 & 14.8329 & -72.1496 & 160.3 & 21.29 & 8.68 & 117.4 & 2.03 & 26823.0 & 7.53 & 32212 & 2345.94 \\ 
        61 & 14.7350 & -72.1742 & 155.7 & 1.69 & 1.24 & 52.0 & 0.71 & 160.4 & 0.80 & 415 & 14.03 \\ 
        62 & 14.7747 & -72.1759 & 159.7 & 5.77 & 3.41 & 111.9 & 2.09 & 2622.1 & 2.46 & 11109 & 229.33\\ 
        63 & 14.8458 & -72.1454 & 155.8 & 1.48 & 0.76 & 89.7 & 0.94 & 15.1 & 0.59 & 539 & 1.32 \\ 
        64 & 14.7343 & -72.1545 & 157.0 & 1.67 & 0.78 & 141.1 & 1.28 & 241.3 & 0.63 & 1075 & 21.10  \\ 
        65 & 14.7668 & -72.1786 & 155.6 & 1.78 & 1.00 & 153.1 & 0.54 & 9.7 & 0.74 & 223 & 0.85  \\ 
        66 & 14.7736 & -72.1910 & 156.1 & 1.56 & 1.03 & -174.5 & 0.69 & 214.3 & 0.70 & 352 & 18.74  \\ 
        67 & 14.7225 & -72.1615 & 158.4 & 10.69 & 2.70 & 67.0 & 1.06 & 2038.7 & 2.97 & 3444 & 178.30 \\ 
        68 & 14.8086 & -72.1499 & 156.0 & 1.79 & 0.96 & -166.6 & 0.55 & 13.5 & 0.73 & 230 & 1.18  \\ 
        69 & 14.7945 & -72.1564 & 157.8 & 2.33 & 1.06 & -145.3 & 0.93 & 23.6 & 0.87 & 786 & 2.06  \\ 
        70 & 14.7073 & -72.1545 & 158.4 & 4.05 & 1.36 & 58.2 & 1.40 & 906.6 & 1.30 & 2640 & 79.29  \\ 
        71 & 14.7477 & -72.1711 & 158.0 & 1.89 & 1.10 & -154.5 & 1.07 & 341.8 & 0.80 & 946 & 29.89  \\ 
        72 & 14.6989 & -72.1563 & 157.0 & 1.51 & 1.09 & 96.9 & 0.53 & 91.8 & 0.71 & 204 & 8.03 \\ 
        73 & 14.8329 & -72.1372 & 158.2 & 1.82 & 1.36 & 159.3 & 0.55 & 193.2 & 0.87 & 277 & 16.90  \\ 
        74 & 14.7610 & -72.1765 & 159.6 & 2.47 & 1.16 & 45.5 & 1.12 & 428.9 & 0.94 & 1231 & 37.51  \\ 
        75 & 14.7896 & -72.1826 & 160.3 & 1.90 & 0.89 & -169.5 & 1.29 & 602.5 & 0.72 & 1241 & 52.69  \\ 
        76 & 14.8003 & -72.1786 & 157.4 & 1.85 & 0.79 & 107.2 & 0.46 & 12.0 & 0.67 & 144 & 1.05  \\ 
        77 & 14.8161 & -72.1637 & 157.2 & 3.06 & 1.05 & 158.2 & 0.39 & 15.4 & 1.00 & 156 & 1.35  \\ 
        78 & 14.7663 & -72.1799 & 158.6 & 1.42 & 0.79 & 160.3 & 0.78 & 61.0 & 0.59 & 372 & 5.34  \\ 
        79 & 14.7161 & -72.1550 & 157.3 & 2.32 & 1.87 & 158.6 & 0.38 & 141.0 & 1.15 & 175 & 12.33  \\ 
        80 & 14.7021 & -72.1645 & 157.5 & 1.43 & 0.88 & -150.1 & 0.36 & 9.6 & 0.62 & 85 & 0.84  \\ 
        81 & 14.7352 & -72.1656 & 159.8 & 3.78 & 1.45 & 143.7 & 1.07 & 885.6 & 1.30 & 1548 & 77.46  \\ 
        82 & 14.8118 & -72.1628 & 158.3 & 1.67 & 0.66 & 160.2 & 0.87 & 14.7 & 0.58 & 463 & 1.28  \\ 
        83 & 14.8504 & -72.1591 & 157.9 & 1.87 & 1.32 & 81.0 & 0.33 & 47.6 & 0.87 & 97 & 4.16  \\ 
        84 & 14.8390 & -72.1622 & 159.2 & 4.91 & 1.88 & 148.0 & 1.05 & 188.9 & 1.68 & 1912 & 16.52 \\ 
        85 & 14.7909 & -72.1517 & 159.6 & 3.01 & 0.99 & 166.1 & 0.86 & 51.5 & 0.96 & 735 & 4.51  \\ 
        86 & 14.7314 & -72.1694 & 160.3 & 2.74 & 0.91 & 52.5 & 0.64 & 156.3 & 0.88 & 379 & 13.67  \\ 
        87 & 14.8599 & -72.1557 & 158.7 & 1.45 & 0.85 & 155.8 & 0.46 & 24.3 & 0.61 & 133 & 2.13  \\ 
        88 & 14.7596 & -72.1727 & 161.3 & 4.81 & 0.93 & -147.6 & 1.43 & 229.5 & 1.17 & 2511 & 20.08 \\ 
        89 & 14.7419 & -72.1734 & 159.7 & 3.66 & 1.44 & 50.8 & 0.92 & 165.5 & 1.27 & 1122 & 14.47  \\ 
        90 & 14.8006 & -72.1615 & 160.9 & 3.08 & 1.07 & 87.5 & 0.52 & 264.4 & 1.00 & 279 & 23.13  \\ 
        91 & 14.8204 & -72.1445 & 160.2 & 3.13 & 1.66 & -163.3 & 0.43 & 143.9 & 1.26 & 237 & 12.59 \\ 
        92 & 14.8516 & -72.1545 & 160.1 & 2.07 & 0.67 & 122.2 & 0.40 & 14.3 & 0.65 & 110 & 1.25 \\ 
        93 & 14.8166 & -72.1600 & 160.6 & 1.03 & 0.97 & -179.8 & 0.51 & 143.3 & 0.55 & 149 & 12.53  \\ 
        94 & 14.8244 & -72.1597 & 161.2 & 1.24 & 0.99 & 105.7 & 0.48 & 56.7 & 0.61 & 147 & 4.96  \\ 
        95 & 14.8051 & -72.1688 & 161.2 & 1.75 & 0.74 & -146.6 & 0.77 & 13.1 & 0.63 & 385 & 1.14  \\ 
        96 & 14.7905 & -72.1677 & 161.6 & 1.43 & 0.73 & 153.3 & 0.79 & 8.1 & 0.57 & 370 & 0.71  \\ 
        97 & 14.8017 & -72.1662 & 164.0 & 1.98 & 1.22 & 145.5 & 1.26 & 534.1 & 0.86 & 1423 & 46.71 \\ 
        98 & 14.7177 & -72.1775 & 162.4 & 1.82 & 0.96 & -136.9 & 0.92 & 16.3 & 0.73 & 645 & 1.42  \\ 
        99 & 14.8405 & -72.1492 & 162.9 & 1.17 & 0.92 & 98.7 & 0.43 & 25.3 & 0.57 & 109 & 2.21  \\ 
        100 & 14.8581 & -72.1428 & 164.3 & 2.04 & 0.92 & 89.5 & 0.90 & 133.4 & 0.76 & 639 & 11.67  \\ 
        101 & 14.8106 & -72.1621 & 163.5 & 1.37 & 1.10 & 157.1 & 0.48 & 83.1 & 0.68 & 166 & 7.27  \\ 
        102 & 14.7934 & -72.1800 & 163.7 & 1.72 & 1.24 & -169.0 & 0.39 & 25.3 & 0.81 & 129 & 2.21  \\ 
        103 & 14.8302 & -72.1463 & 164.1 & 2.00 & 1.31 & 108.2 & 0.42 & 74.1 & 0.90 & 167 & 6.48  \\ 
        104 & 14.7965 & -72.1681 & 167.6 & 1.39 & 0.97 & 153.4 & 0.61 & 19.7 & 0.64 & 252 & 1.73 \\
\label{tab:Trunks}

\end{longtable}

\begin{longtable}{lllllllllllll}
\caption[] {\textsuperscript{12}CO(1–0) Core (Leave) Properties.\\
}\\

\hline
ID  &  RA      & Dec.     & $v_{\mathrm{cen}}$  & $\sigma_{\mathrm{maj}}$   & $\sigma_{\mathrm{min}}$  & Angle \textsuperscript{1} & $v_{\mathrm{rms}} $ & Flux \textsuperscript{2}  & Radius  & $M_{\mathrm{vir}}$  & $L_{\mathrm{CO}}$   \\
   & (deg)     & (deg)     & (km s$^{-1}$)  & (arcsec) &  (arcsec) &  (deg) & (km s$^{-1}$) & ( K ) & ( pc ) &  ($\mathrm{M_{\odot}}$) &  (K km s$^{-1}$ pc$^2$)  \\ \hline
\endfirsthead 
 \caption[]{(Cores properties continued)}\\
 
\hline
ID  &  RA      & Dec.     & $v_{\mathrm{cen}}$  & $\sigma_{\mathrm{major}}$  & $\sigma_{\mathrm{minor}}$  & Angle \textsuperscript{1}  & $v_{\mathrm{rms}} $ & Flux \textsuperscript{2}  & Radius  & $M_{\mathrm{vir}}$  & $L_{\mathrm{CO}}$  \\
   & (deg)     & (deg)     & (km s$^{-1}$)  & (arcsec) &  (arcsec) &  (deg) & (km s$^{-1}$) & ( K ) & ( pc ) &  ($\mathrm{M_{\odot}}$) &  (K km s$^{-1}$ pc$^2$)   \\ \hline
\endhead
\hline 
\multicolumn{12}{l}{\textsuperscript{1} The orientation angle of the major axis.} \\
\multicolumn{12}{l}{\textsuperscript{2} The integrated flux of a structure.}
\endlastfoot
   
         1 & 14.7357 & -72.1818 & 139.2 & 2.31 & 1.11 & -171.6 & 1.02 & 60.1 & 0.89 & 962 & 5.26 \\ 
        2 & 14.7384 & -72.1838 & 144.1 & 2.71 & 0.89 & -136.7 & 0.78 & 102.4 & 0.86 & 548 & 8.96 \\ 
        3 & 14.7697 & -72.1846 & 145.2 & 3.89 & 0.89 & 172.7 & 1.19 & 1027.4 & 1.03 & 1530 & 89.86  \\ 
        4 & 14.7253 & -72.1797 & 144.2 & 1.53 & 1.28 & 54.2 & 0.67 & 150.1 & 0.78 & 358 & 13.13  \\ 
        5 & 14.7205 & -72.1783 & 143.9 & 1.49 & 0.90 & -178.7 & 0.63 & 140.3 & 0.64 & 264 & 12.27 \\ 
        6 & 14.7166 & -72.1762 & 145.9 & 1.79 & 0.81 & 48.1 & 0.81 & 80.6 & 0.67 & 451 & 7.05  \\ 
        7 & 14.8307 & -72.1872 & 147.5 & 2.01 & 1.07 & 69.3 & 0.90 & 163.9 & 0.81 & 686 & 14.33  \\ 
        8 & 14.7174 & -72.1785 & 146.8 & 1.74 & 1.25 & -172.5 & 0.57 & 53.0 & 0.82 & 272 & 4.63  \\ 
        9 & 14.7165 & -72.1672 & 147.6 & 1.54 & 0.87 & 131.5 & 0.60 & 83.7 & 0.64 & 241 & 7.32  \\ 
        10 & 14.8345 & -72.1894 & 151.5 & 2.91 & 1.63 & 51.1 & 0.97 & 2164.9 & 1.21 & 1191 & 189.34 \\ 
        11 & 14.7143 & -72.1711 & 151.0 & 1.68 & 0.62 & 140.5 & 0.82 & 193.0 & 0.57 & 399 & 16.88  \\ 
        12 & 14.8016 & -72.2037 & 151.7 & 1.82 & 0.90 & 164.2 & 0.66 & 97.2 & 0.71 & 326 & 8.50  \\ 
        13 & 14.7395 & -72.1657 & 152.1 & 2.51 & 1.25 & 59.0 & 0.66 & 193.0 & 0.98 & 440 & 16.88  \\ 
        14 & 14.8089 & -72.2023 & 151.6 & 1.57 & 1.39 & 61.5 & 0.45 & 241.9 & 0.82 & 173 & 21.16 \\ 
        15 & 14.7181 & -72.1785 & 151.8 & 2.13 & 0.77 & -160.5 & 0.48 & 15.6 & 0.71 & 172 & 1.36 \\ 
        16 & 14.8273 & -72.1903 & 153.8 & 3.16 & 1.07 & -170.4 & 0.80 & 433.6 & 1.02 & 679 & 37.93 \\ 
        17 & 14.7253 & -72.1676 & 154.3 & 2.27 & 1.00 & 100.4 & 0.69 & 438.5 & 0.83 & 417 & 38.35 \\ 
        18 & 14.8174 & -72.1914 & 155.4 & 2.16 & 0.54 & 159.6 & 0.41 & 116.7 & 0.60 & 105 & 10.21 \\ 
        19 & 14.7282 & -72.1674 & 155.9 & 1.27 & 0.82 & 53.2 & 0.53 & 156.7 & 0.57 & 164 & 13.70 \\ 
        20 & 14.7722 & -72.1745 & 156.9 & 1.89 & 1.00 & -162.6 & 0.61 & 345.3 & 0.76 & 294 & 30.20 \\ 
        21 & 14.7080 & -72.1548 & 157.3 & 1.94 & 1.37 & 84.6 & 0.59 & 422.4 & 0.90 & 322 & 36.94 \\ 
        22 & 14.8359 & -72.1499 & 157.3 & 4.29 & 0.83 & 140.2 & 0.73 & 559.3 & 1.05 & 574 & 48.92 \\ 
        23 & 14.8362 & -72.1627 & 159.0 & 2.28 & 1.32 & 165.8 & 0.91 & 103.4 & 0.96 & 835 & 9.04 \\ 
        24 & 14.8255 & -72.1506 & 158.1 & 3.09 & 0.93 & 84.0 & 0.72 & 265.5 & 0.94 & 507 & 23.22 \\ 
        25 & 14.7260 & -72.1640 & 158.7 & 1.39 & 1.31 & -154.8 & 0.81 & 616.0 & 0.75 & 511 & 53.88 \\ 
        26 & 14.8208 & -72.1539 & 159.3 & 2.66 & 1.94 & 154.9 & 0.99 & 1864.6 & 1.26 & 1294 & 163.08\\ 
        27 & 14.7182 & -72.1595 & 158.1 & 2.90 & 0.50 & 64.8 & 0.52 & 133.4 & 0.67 & 186 & 11.67  \\ 
        28 & 14.8431 & -72.1614 & 159.2 & 2.27 & 1.25 & -135.3 & 0.88 & 78.0 & 0.94 & 749 & 6.82  \\ 
        29 & 14.8487 & -72.1364 & 158.9 & 1.51 & 0.66 & 150.4 & 0.54 & 141.9 & 0.55 & 169 & 12.41  \\ 
        30 & 14.8058 & -72.1581 & 159.6 & 1.95 & 0.75 & 80.2 & 0.67 & 123.4 & 0.67 & 313 & 10.79  \\ 
        31 & 14.7425 & -72.1741 & 159.4 & 1.34 & 0.83 & 157.8 & 0.86 & 44.8 & 0.59 & 446 & 3.92  \\ 
        32 & 14.7235 & -72.1622 & 159.1 & 1.42 & 0.74 & 115.8 & 0.51 & 187.7 & 0.57 & 157 & 16.42 \\ 
        33 & 14.7746 & -72.1769 & 160.7 & 3.17 & 1.18 & 111.5 & 0.76 & 1062.1 & 1.07 & 636 & 92.90 \\ 
        34 & 14.7403 & -72.1727 & 159.5 & 1.56 & 0.75 & 63.9 & 0.68 & 85.6 & 0.60 & 290 & 7.48  \\ 
        35 & 14.7311 & -72.1663 & 159.6 & 1.58 & 0.73 & 149.6 & 0.74 & 147.0 & 0.60 & 336 & 12.85 \\ 
        36 & 14.8440 & -72.1385 & 159.8 & 2.90 & 1.81 & 121.6 & 0.57 & 474.6 & 1.27 & 435 & 41.51  \\ 
        37 & 14.8289 & -72.1522 & 159.5 & 1.61 & 1.16 & 141.4 & 0.52 & 662.9 & 0.76 & 214 & 57.98 \\ 
        38 & 14.8380 & -72.1444 & 159.7 & 1.39 & 0.92 & 90.7 & 0.56 & 501.6 & 0.63 & 203 & 43.87 \\ 
        39 & 14.8360 & -72.1407 & 160.1 & 1.19 & 0.93 & 102.5 & 0.66 & 232.5 & 0.58 & 262 & 20.33 \\ 
        40 & 14.7054 & -72.1537 & 159.7 & 3.50 & 0.72 & 54.1 & 0.37 & 251.8 & 0.88 & 128 & 22.02  \\ 
        41 & 14.8101 & -72.1524 & 160.0 & 3.91 & 1.21 & 115.8 & 0.52 & 126.7 & 1.21 & 332 & 11.08 \\ 
        42 & 14.8226 & -72.1447 & 160.1 & 1.83 & 0.66 & 118.9 & 0.41 & 64.2 & 0.61 & 104 & 5.61 \\ 
        43 & 14.8488 & -72.1463 & 160.7 & 1.90 & 0.89 & 141.6 & 0.49 & 513.6 & 0.72 & 182 & 44.92 \\ 
        44 & 14.8347 & -72.1463 & 160.4 & 1.52 & 1.16 & 102.4 & 0.40 & 320.6 & 0.73 & 122 & 28.04 \\ 
        45 & 14.8178 & -72.1443 & 160.3 & 2.11 & 1.10 & -135.2 & 0.31 & 55.9 & 0.84 & 84 & 4.89 \\ 
        46 & 14.7105 & -72.1561 & 160.1 & 1.65 & 0.64 & 84.7 & 0.25 & 51.9 & 0.57 & 36 & 4.54  \\ 
        47 & 14.7614 & -72.1730 & 162.0 & 1.52 & 0.71 & -150.7 & 0.78 & 61.1 & 0.58 & 367 & 5.35 \\ 
        48 & 14.8141 & -72.1499 & 160.8 & 1.97 & 1.34 & 141.7 & 0.30 & 95.2 & 0.90 & 86 & 8.33 \\ 
        49 & 14.8465 & -72.1502 & 163.6 & 3.82 & 1.21 & 97.8 & 0.92 & 497.0 & 1.19 & 1045 & 43.46 \\ 
        50 & 14.8214 & -72.1477 & 161.5 & 1.30 & 1.03 & -170.5 & 0.29 & 59.8 & 0.64 & 55 & 5.23 \\ 
        51 & 14.8173 & -72.1576 & 162.3 & 2.05 & 0.81 & 87.0 & 0.56 & 94.1 & 0.71 & 232 & 8.23 \\ 
        52 & 14.8330 & -72.1534 & 161.8 & 1.94 & 0.55 & -161.5 & 0.23 & 97.9 & 0.57 & 32 & 8.56 \\ 
        53 & 14.8537 & -72.1464 & 162.0 & 1.33 & 1.12 & 69.3 & 0.32 & 121.7 & 0.68 & 72 & 10.64 \\ 
        54 & 14.7799 & -72.1735 & 162.2 & 2.47 & 0.69 & 134.1 & 0.34 & 66.1 & 0.73 & 86 & 5.78 \\ 
        55 & 14.8399 & -72.1531 & 162.0 & 1.68 & 0.74 & 88.9 & 0.35 & 194.5 & 0.62 & 79 & 17.01 \\ 
        56 & 14.8475 & -72.1416 & 163.8 & 3.11 & 1.23 & 144.0 & 0.75 & 119.2 & 1.08 & 631 & 10.42 \\ 
        57 & 14.8303 & -72.1558 & 162.9 & 3.02 & 1.04 & 114.6 & 0.44 & 775.3 & 0.98 & 200 & 67.80 \\ 
        58 & 14.8427 & -72.1540 & 162.9 & 1.54 & 0.75 & 55.0 & 0.31 & 186.8 & 0.60 & 59 & 16.34 \\ 
        59 & 14.8247 & -72.1544 & 163.6 & 1.70 & 0.74 & 143.9 & 0.49 & 122.9 & 0.62 & 156 & 10.75 \\ 
        60 & 14.8522 & -72.1485 & 164.1 & 2.24 & 0.84 & 111.6 & 0.32 & 49.8 & 0.76 & 80 & 4.36 \\ 
        61 & 14.8236 & -72.1572 & 164.4 & 1.96 & 0.80 & 149.4 & 0.40 & 60.8 & 0.69 & 117 & 5.32 \\ 
\label{tab:leaves}
\end{longtable}




\bibliography{sample631}{}
\bibliographystyle{aasjournal}



\end{document}